\documentclass[preprint,12pt,authoryear,review]{elsarticle}
\usepackage{graphicx} 
\usepackage{lineno}
\usepackage[hyphens]{url}
\usepackage{hyperref}
\usepackage{etoolbox}
\apptocmd{\thebibliography}{\raggedright}{}{}
\usepackage{siunitx} 
\usepackage{subcaption}
\usepackage{amssymb}
\usepackage{braket}
\usepackage{amsmath}
\usepackage{color}
\usepackage{makecell}

\usepackage{adjustbox}

\usepackage{xcolor}

\journal{EPSL}

\begin{document}

\begin{frontmatter}

\title{Sensitivity of neutrino oscillations to the Earth's interior properties}

\author[label1,label2]{Isabel Goos}
\author[label2,label3]{Nobuaki Fuji}
\author[label4]{Stéphanie Durand}
\author[label1,label3]{V\'eronique Van Elewyck}
\author[label1]{João A. B. Coelho}
\author[label5]{Eric Mittelstaedt}
\author[label5]{Yael Deniz}

\affiliation[label1]{organization={Université Paris Cité, CNRS, Astroparticule et Cosmologie},
            city={Paris},
            postcode={F-75013}, 
            country={France}}

\affiliation[label2]{organization={Université Paris Cité, Institut de physique du globe de Paris, CNRS},
            city={Paris},
            postcode={F-75005}, 
            country={France}}

\affiliation[label3]{organization={Institut universitaire de France},
            city={Paris},
            country={France}}

\affiliation[label4]{
    organization={Université Claude Bernard Lyon 1, ENS de Lyon, CNRS, UMR 5276 LGL-TPE},
    city={Villeurbanne},
    country={France}}

\affiliation[label5]{
            organization={University of Idaho, Department of Earth and Spatial Sciences},
            city={Moscow},
            postcode={83844}, 
            state={Idaho},
            country={United States of America}}

\begin{abstract}
Understanding the Earth’s deep interior remains challenging, as traditional geophysical methods face ambiguities in linking seismic observations to temperature, composition, or mass density variations. Atmospheric neutrinos, produced by the constant flux of cosmic rays colliding with the upper atmosphere, offer a complementary probe: with energies of a few GeV, they traverse the Earth and experience flavor oscillations influenced by the planet’s electron density distribution, which depends on both its mass density and composition. Combining seismic observations with neutrino measurements in a joint inversion framework could provide complementary constraints on gross spatial and compositional variations in the deep Earth. We consider the next-generation neutrino detectors KM3NeT/ORCA, Hyper-Kamiokande, and DUNE, as well as an idealized hypothetical detector representing the intrinsic sensitivity of the method. The idealized detector is most sensitive to perturbations in the core, whereas realistic detector performance shifts the sensitivity toward the mantle, including the mantle transition zone, where hydrogen enrichment could produce detectable variations in electron density. These results establish the sensitivity of neutrino oscillations to the Earth’s electron density profile and provide a basis for future joint neutrino–seismic tomography.
\end{abstract}

\begin{graphicalabstract}
\begin{figure}[htbp]
    \begin{center}
    \includegraphics[width=\textwidth]{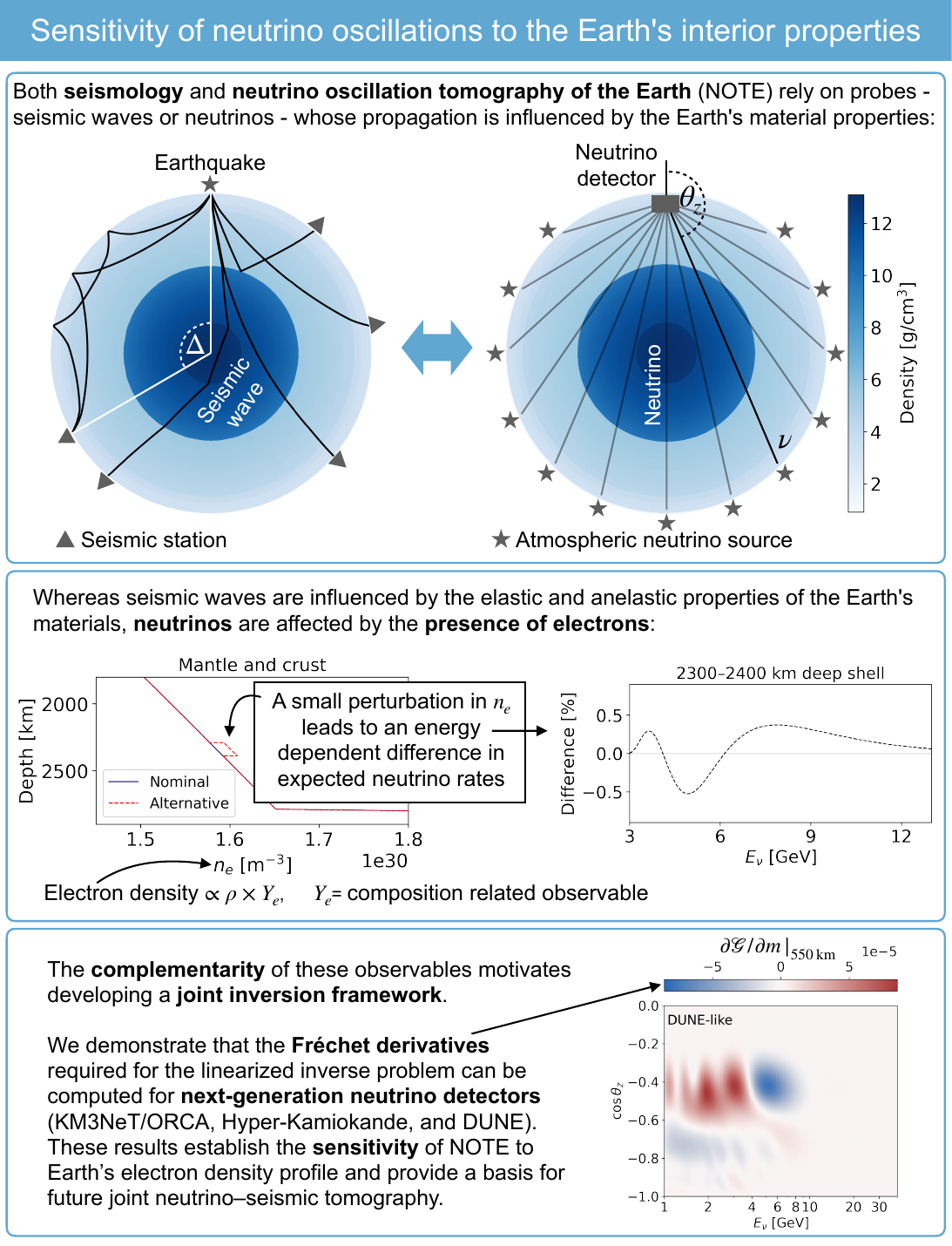}
    \end{center}
    \label{fig:nutomoVSseismo}
\end{figure}
\end{graphicalabstract}

\begin{highlights}
\item Interdisciplinary framework linking neutrino physics and Earth sciences
\item Pedagogical forward modeling of neutrino oscillation tomography
\item Sensitivity kernels for inverting for electron density
\item Framework for a future neutrino and seismic data joint inversion 
\end{highlights}
\begin{keyword}
   Earth’s interior \sep atmospheric neutrinos \sep neutrino oscillations \sep density profile inversion \sep seismology
\end{keyword}

\end{frontmatter}

\section{Introduction}
\label{sec:intro}

Direct exploration of the Earth's interior is limited to depths of approximately \SI{12}{\kilo\m}, corresponding to the deepest borehole ever drilled~\citep{kozlovsky1987superdeep}. Consequently, understanding the planet’s interior has relied on indirect methods. Among these, seismology~\citep{aki1976three, sengupta1976three, fichtner2024seismic} has been particularly successful in revealing its internal \mbox{3D} structure through inversion of seismic data. Seismologists use either full seismic waveforms~\citep{fichtner2010full} or their secondary attributes such as the travel times of body waves~\citep{aki1976three, sengupta1976three,hosseini2020global} and the dispersion of surface waves~\citep{montagner1990global,durand2016seismic,debayle2016}, or a combination of these~\citep{durand2016,durand2017,ritsema2011,koelemeijer2016}. Both approaches constrain seismological parameters, such as isotropic and anisotropic seismic velocities, and attenuation, which are then used to infer physical properties of the Earth's interior, such as temperature, pressure, and chemical composition. Because the sensitivity of body waves and surface waves to density is much lower than to seismic velocities, imaging density variations inside Earth has proven difficult. Free oscillations excited by large earthquakes (Mw $\geq 7.0$) provide one of the few geophysical tools for decoupling elastic moduli and density with sufficiently strong sensitivity. This approach has helped constrain the radial density profile~\citep{dziewonski1981preliminary,irving2018}, but inferring a full \mbox{3D} density structure remains challenging~\citep{ishii1999,kuo2002,rom2001}.

Given the expansion of seismic networks, increased computational resources, and improvements in direct and inverse problem theory, the resolution of global \mbox{3D} tomographic models has improved continuously. However, while seismic tomography effectively images structure in terms of seismic properties, inverting for temperature and composition from seismic data alone is an ill-posed problem that requires extensive assumptions and extrapolations of mineral physics, petrological, and thermochemical data~\citep{deng2023compositional}. To address these limitations, combining multiple complementary observables, each sensitive to different physical properties, offers an alternative way to better constrain the origin of seismic heterogeneities. 

In this context, neutrino oscillation tomography of the Earth (NOTE) has recently emerged as a promising complementary approach~\citep{winter2006nutomo}. NOTE offers the possibility to probe the Earth's interior using neutrinos, naturally abundant fundamental particles that traverse the planet~\citep{athar2022status}. NOTE makes use of atmospheric neutrinos, which are copiously produced in extensive air showers generated by the interaction of cosmic rays with air molecules \citep[Chapter 24 of~][]{tanabashi2018review}. Neutrinos come in three types, known as flavors: electron neutrino $\nu_e$, muon neutrino $\nu_\mu$, and tau neutrino $\nu_\tau$, together with their antiparticles $\bar{\nu}_e$, $\bar{\nu}_\mu$, and $\bar{\nu}_\tau$. Once produced in the atmosphere, they propagate in straight lines through the Earth. As they propagate, neutrinos have a probability to change from one flavor to another, a quantum-mechanical phenomenon known as flavor oscillations. Consequently, a neutrino produced in a given flavor could be detected after some distance in a different flavor. The presence of electrons (i.e., matter) in the Earth affects these oscillation probabilities for neutrinos with energies from a few GeV to several tens of GeV. NOTE exploits this phenomenon to probe the Earth’s electron-density profile, which combines matter density and chemical composition, by analyzing distortions in oscillation patterns~\citep{rott2015spectrometry,winter2016atmospheric,VanElewyck:20174h,Denton:2021rgt,DOlivoSaez:2022vdl,Upadhyay:2022jfd,maderer2023unveiling,petcov2024neutrino}.

NOTE has not yet been performed using experimental data. The neutrino detectors Hyper-Kamiokande (HK) and DUNE are under construction, while KM3NeT/ORCA has accumulated only a limited amount of data. Consequently, current results are limited to sensitivity projections. For DUNE, $10$~years of measurements are expected to yield sensitivities of $\sim$10\% for the matter density of the core and lower mantle, and $\sim$20\% for the upper mantle~\citep{kelly2022dune}. HK is expected to constrain the average core density at the $\mathcal{O}(10\text{–}40\%)$ level after $\sim$20~years of measurements, depending on the achieved measurement performance. While not competitive with geophysical constraints, NOTE can provide direct sensitivity to chemical composition. For instance, ~\citet{maderer2023unveiling} show that combining KM3NeT/ORCA, HK, and DUNE could discriminate FeNi from FeNiH outer-core compositions at $1\sigma$ after $\sim$50~years of concomitant measurements. Neutrino measurements, while individually less constraining, could help break degeneracies inherent to geophysical probes. Finally, simulations based on the neutrino detector IceCube DeepCore, representing $\sim$10~years of operation, demonstrate the ability to perform correlated density measurements of different Earth layers when incorporating constraints from the Earth’s total mass and moment of inertia~\citep{chattopadhyay2025demonstrating}.

In this study, we propose to exploit neutrino observables to constrain Earth’s interior properties and aim to lay the groundwork for an inverse problem based on neutrino data. Solving any inverse problem requires defining the forward problem, in which the synthetic data $\mathbf{u}$, the model parameters $\mathbf{m}$, and the forward function $\mathcal{G}$ are characterized such that $\mathbf{u}=\mathcal{G}(\mathbf{m})$. In many cases progress toward inversion relies on a linearized description of this function through the Fréchet derivatives $\nabla_\mathbf{m}\mathcal{G}(\mathbf{m})$. In this work, we establish the mathematical framework required for neutrino oscillation based inversion and derive the associated Fréchet derivatives. We then perform a comprehensive sensitivity study of NOTE with respect to variations in Earth’s physical properties. These results provide the necessary basis for a future joint inversion with seismic data.

The paper is organized as follows. After the introduction in Section~\ref{sec:intro}, Section~\ref{sec:nutomoVSseismo} contrasts neutrino and seismic observables. Section~\ref{sec:forward} introduces the forward problem $\mathcal{G}$ for NOTE. Section~\ref{sec:methods} presents the methods used in this work. The computation of Fréchet derivatives is demonstrated in Section~\ref{sec:results}, together with the presentation of the sensitivity study. We conclude with closing remarks in Section~\ref{sec:conclusions}.

\section{Neutrino oscillation tomography in comparison with seismology}
\label{sec:nutomoVSseismo}

This paper aims to lay the groundwork for the future joint analysis of seismic and neutrino data. Before proceeding, it is important to describe neutrino oscillation tomography of the Earth (NOTE) in comparison with seismology, as illustrated in Figure~\ref{fig:nutomoVSseismo}. In both approaches, a probe (neutrino or wave) propagates through the Earth and is influenced by the Earth's material properties along its ray. Neutrinos are affected by the presence of electrons, whereas seismic waves are influenced by the elastic and anelastic properties of the Earth's materials. Upon detection after traversing the Earth, their observed properties reflect the characteristics of the material encountered along their ray. Hence, seismology and NOTE share a common foundation, which motivates our preparatory work toward a joint inversion. A closer look reveals important differences alongside these similarities, as discussed below.

\begin{figure}[htbp]
    \vspace{-2.7cm}
    \begin{center}
    \includegraphics[width=\textwidth]{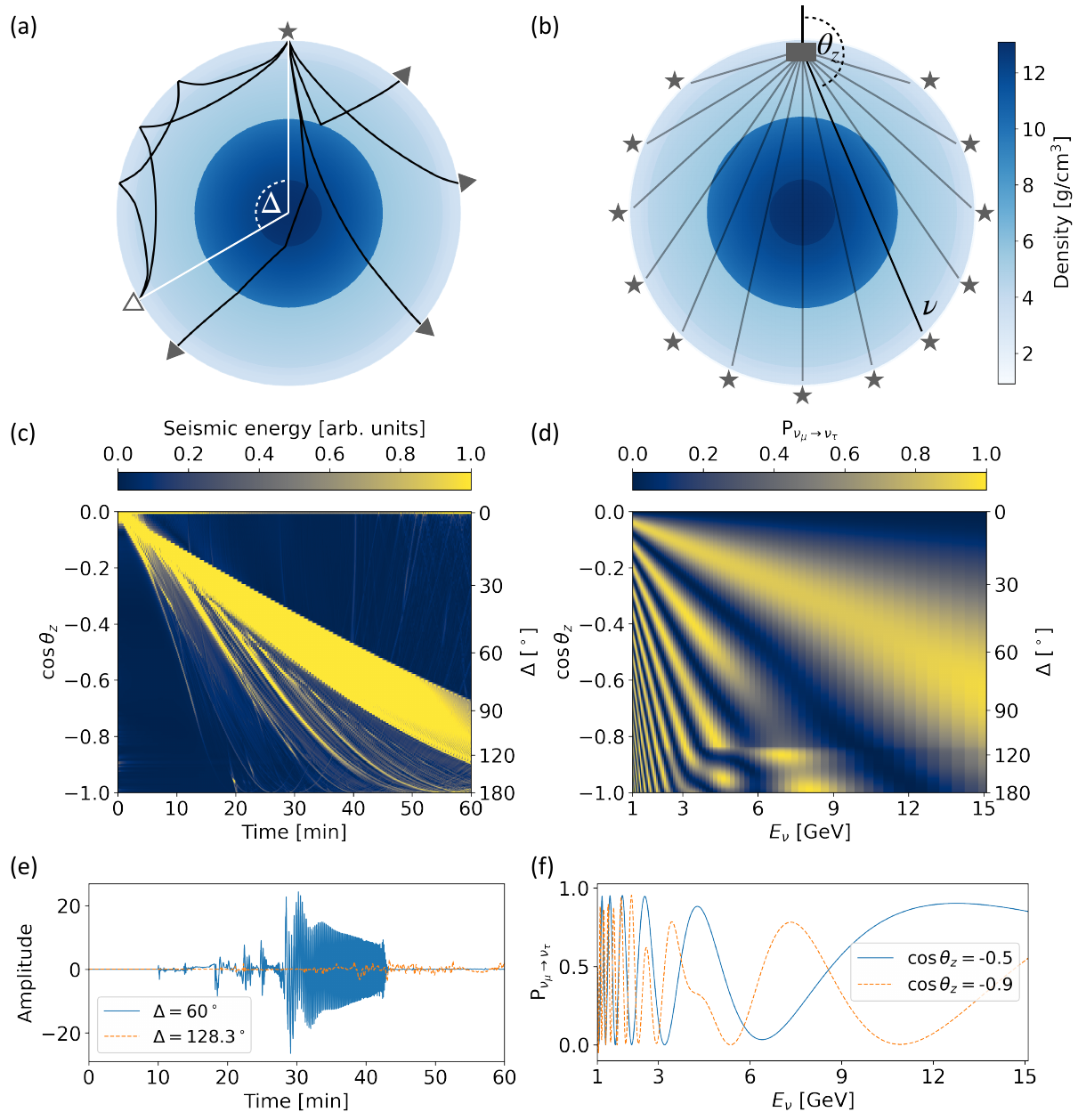}
    \end{center}
    \caption{Overview of the data-acquisition configurations (a-b) and recorded observations (c-f) in both global seismology (left column) and NOTE (right column). (a) Triangles represent seismic stations and the star an earthquake, with rays illustrating seismic wave paths sampling different regions of the Earth. (b) The grey rectangle denotes a neutrino detector and the stars neutrino sources, where the incidence angle $\theta_z$ is measured from the vertical at the detector and straight lines show possible neutrino rays. (c) Time–distance plot obtained by ordering the vertical components of synthetic seismograms by epicentral distance. The synthetic seismograms are calculated using the PREM model~\citep{dziewonski1981preliminary}  for an earthquake at \SI{30}{\kilo\m} depth. We plot the envelopes of the seismograms, representing the energy arriving over time. The time is given with respect to the time of the earthquake. (d) Neutrino oscillogram representing the probability P$_{\nu_\mu \rightarrow \nu_\tau}$ as a function of neutrino energy $E_\nu$ and incidence angle $\theta_z$. The Earth model described in Section~\ref{subsec:nuearth} is used. (e) Seismic waveforms as a function of time for two particular epicentral distances $\Delta$, $60^\circ$ (blue curve) and 128.3$^\circ$ (dashed orange curve). (f) Oscillation probabilities as a function of energy for  two particular incidence angles $\theta_z$, equivalent to $\Delta=60^\circ$ (blue curve) and $\Delta=128.3^\circ$ (dashed orange curve), as in panel (e).}
    \label{fig:nutomoVSseismo}
\end{figure}

First, the data-acquisition configurations of NOTE and terrestrial global seismology are fundamentally inverted (see Figures~\ref{fig:nutomoVSseismo}a and~\ref{fig:nutomoVSseismo}b). Because of their cost, only a limited number of neutrino detectors will be deployed worldwide. Atmospheric neutrinos are produced continuously and abundantly in the Earth's atmosphere, allowing each detector to record neutrinos arriving from all directions. In contrast, earthquakes occur sporadically, whereas seismic stations are distributed globally—albeit unevenly. Additionally, seismology benefits from over five decades of accumulated observations, giving it far greater statistical power than current atmospheric neutrino data. However, next-generation neutrino detectors are expected to substantially increase the NOTE dataset in the coming decades.


Another distinction lies in the traceability of individual signals. In seismology, specific seismic phases recorded on a seismogram can be associated with well-defined propagation rays back to a specific source (Figure~\ref{fig:nutomoVSseismo}a). By contrast, the continuous and intense nature of the atmospheric neutrino flux means that an individual neutrino cannot be traced back to the shower that produced it (Figure~\ref{fig:nutomoVSseismo}b). Rather, for NOTE, the flux as a whole must be accurately modeled.

Another difference concerns seismic wave propagation. Seismic waves undergo reflection and refraction within the Earth's interior, leading to complex ray trajectories (Figure~\ref{fig:nutomoVSseismo}a). Neutrinos propagate along straight rays (Figure~\ref{fig:nutomoVSseismo}b). Consequently, once the incidence angle of a neutrino is measured, its ray, and thus the region of the Earth it probes, is uniquely constrained within uncertainties.


NOTE is based on the principle that the presence of electrons along the neutrino ray impacts its oscillation probabilities. The probability of a given oscillation process $\nu_\alpha \rightarrow \nu_\beta$ (where $\alpha, \beta \in \{ e, \mu, \tau \}$) is usually visualized in a \mbox{2D} plot, called oscillogram. The oscillogram is a function of neutrino energy $E_\nu$ and incidence angle $\theta_z$, where $\theta_z$ is measured with respect to the vertical at the neutrino's exit point on the Earth's surface (Figure~\ref{fig:nutomoVSseismo}b). Figure~\ref{fig:nutomoVSseismo}d displays the oscillogram corresponding to the probability P$_{\nu_\mu \rightarrow \nu_\tau}$ that a muon neutrino produced in the atmosphere interacts as a tau neutrino after having propagated across the Earth. Oscillograms illustrate how oscillations transform the flavor composition of the neutrino flux traversing the Earth, capturing solely the influence of Earth's matter on oscillation probabilities, independent of the neutrino source, and subsequent detection process. They can be compared against time-distance plots in seismology (Figure~\ref{fig:nutomoVSseismo}c). Here, seismograms are ordered as a function of distance from the source and, instead of showing the amplitudes of the seismograms, we display their envelopes, which represent the energy.

In Figure~\ref{fig:nutomoVSseismo}d, a sudden change in the oscillation pattern occurs at an epicentral distance of $\Delta \simeq 114^\circ$, or equivalently $\cos\theta_z \simeq -0.84$. This is due to the large jump in electron density experienced by neutrinos that sample the core compared to those that do not. The envelopes of the seismograms in Figure~\ref{fig:nutomoVSseismo}c also exhibit a change at around $110^\circ$, where the first arrival fades. This feature, though less striking than for neutrinos, is likewise attributable to the presence of the core.

Traveltime tomography extracts first breaks or major phase onsets from seismic waveforms for numerous source–receiver pairs. Setting aside the treatment of Fresnel zones (ray theory versus finite-frequency tomography), each travel time is a scalar quantity that encodes the integral of the structure sampled along the corresponding ray. While a single travel time cannot uniquely constrain the seismic wave velocity in every voxel, the ensemble of intersecting rays enables imaging of Earth's interior. Full waveform inversion goes even further by exploiting complete time-dependent signals~\citep{deschamps2019radial, fuji2010methodology, konishi2020three, French_2015, Virieux2009}, which carry information from time-dependent Fresnel zones, thereby providing substantially more information about the propagation medium. Example waveforms are shown in Figure~\ref{fig:nutomoVSseismo}e for $\Delta = 60^\circ$ and $80^\circ$, corresponding to $\cos\theta_z = -0.5$ and $-0.9$, respectively. 

In NOTE, when considering a single detector, neutrino rays never intersect because they travel in straight lines and their Fresnel zone is restricted to the ray itself (Figure~\ref{fig:nutomoVSseismo}b). An analogy can be drawn with one-station seismology, such as the InSight SEIS experiment on Mars~\citep{Lognonne_2020, Jacob_2022}. By exploiting full waveforms whose sensitivity to structure evolves over time it became possible to image the planet's interior even without crossing rays. Oscillation probabilities exhibit a characteristic dependence on energy (Figure~\ref{fig:nutomoVSseismo}f). More importantly, the impact of perturbations in the matter properties on this probability is also energy dependent. Hence, in principle, imaging a thin column of matter using neutrinos of varying energies that traverse it is feasible, and the prospect of imaging the entire Earth using neutrinos that sample different regions motivates the present work.

\section{Forward modeling of neutrino observables}
\label{sec:forward}

\begin{figure}[!htb]
    \centering
    \vspace{-3.5cm}
    \hspace{2.2cm}
    \includegraphics[width=0.8\linewidth]{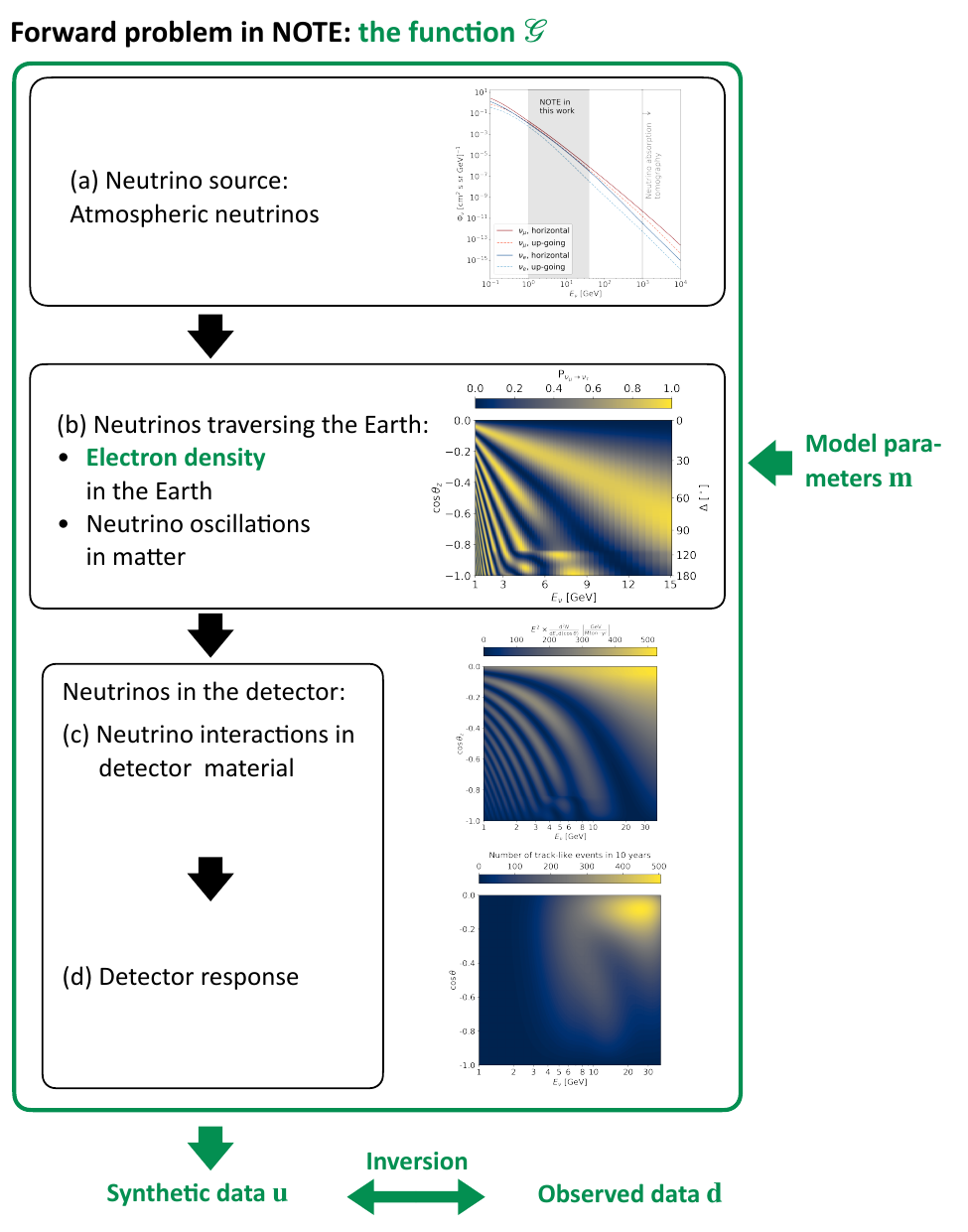}
    \caption{Flowchart of the forward modeling of neutrino observables through the function $\mathcal{G}$. The process starts with the atmospheric neutrino flux (step (a)), as described in Section~\ref{subsec:nuflux}, using the  model from~\citet{honda2015atmospheric}, with the corresponding Figure~\ref{fig:honda} included here in miniature for illustration. Atmospheric neutrinos then traverse the Earth, parameterized by the model parameters $\mathbf{m}$, and undergo flavor transitions (step (b)), whose probabilities depend on the electron density along their ray~\citep{dziewonski1981preliminary, mcdonough2010compositional, wedepohl1995composition, quinby1983distribution} as well as on neutrino oscillation physics in matter~\citep{giunti2007fundamentals}, as outlined in Section~\ref{subsec:nuearth}. The example oscillogram from Figure~\ref{fig:nutomoVSseismo}d is also embedded. The figures shown within the boxes in steps (a) and (b) correspond directly to these processes (the oscillogram represents oscillation probabilities only and is independent of the neutrino flux). In the vicinity of the detector, neutrinos may interact (step (c)) and produce detectable signals (step (d)), as described in Section~\ref{subsec:nudet}~\citep{formaggio2012ev, adrian2016letter, abe2018hyper, abi2020volume}. Figures~\ref{fig:IntsDets}a and~\ref{fig:IntsDets}b correspond to the result of coupling steps (a)–(c) and (a)–(d), respectively, and are shown outside the boxes to highlight this distinction from the figures in (a) and (b). The outcome of combining steps (a)–(d) is the synthetic data $\mathbf{u}$, which can be compared with real measurements $\mathbf{d}$ to perform an inversion.}
    \label{fig:g}
\end{figure}

In this section, we outline the forward problem of NOTE, represented by the function $\mathcal{G}$, which maps Earth properties $\mathbf{m}$ to expected neutrino rates, denoted $\mathbf{u}$. These synthetic data can be compared with observed data $\mathbf{d}$ to perform an inversion. Figure~\ref{fig:g} summarizes the three main components of $\mathcal{G}$: the flux of atmospheric neutrinos entering the Earth (Section~\ref{subsec:nuflux}), the computation of oscillation probabilities as they traverse the Earth's interior (Section~\ref{subsec:nuearth}), and the modeling of their detection (Section~\ref{subsec:nudet}). For reference, ~\ref{sec:tables} provides tables listing the physical constants, variables, and expressions used in this study.

\subsection{Atmospheric neutrino flux}
\label{subsec:nuflux}

Atmospheric neutrinos are produced in the atmosphere when cosmic rays interact with air molecules (step (a) in Figure~\ref{fig:g}). The cosmic ray spectrum comprises nuclei ranging from hydrogen to iron, spanning energies from a few hundred MeV to $\sim$\SI{e14}{\mega\eV}. The average altitude at which cosmic rays first interact with the atmosphere is nucleus-dependent. Hydrogen interacts on average at around \SI{16}{\kilo\m}~\citep{blanco2009cosmic}, while heavier nuclei interact at progressively higher altitudes. Other sources of shower-to-shower fluctuations, as well as atmospheric density variations, further influence neutrino production.

In this study, we use the atmospheric neutrino flux model by~\citet{honda2015atmospheric}. It simulates the cosmic ray flux and its interactions with the atmosphere, as well as the propagation, interactions, and decays of secondary particles, while accounting for all the sources of variability mentioned above. It provides neutrino fluxes as a function of energy and arrival direction at several reference experimental sites. The specific configuration used in this study is detailed in~\ref{subsec:nuflux2}.

\begin{figure}[!htb]
    \centering
    \includegraphics[width=0.7\linewidth]{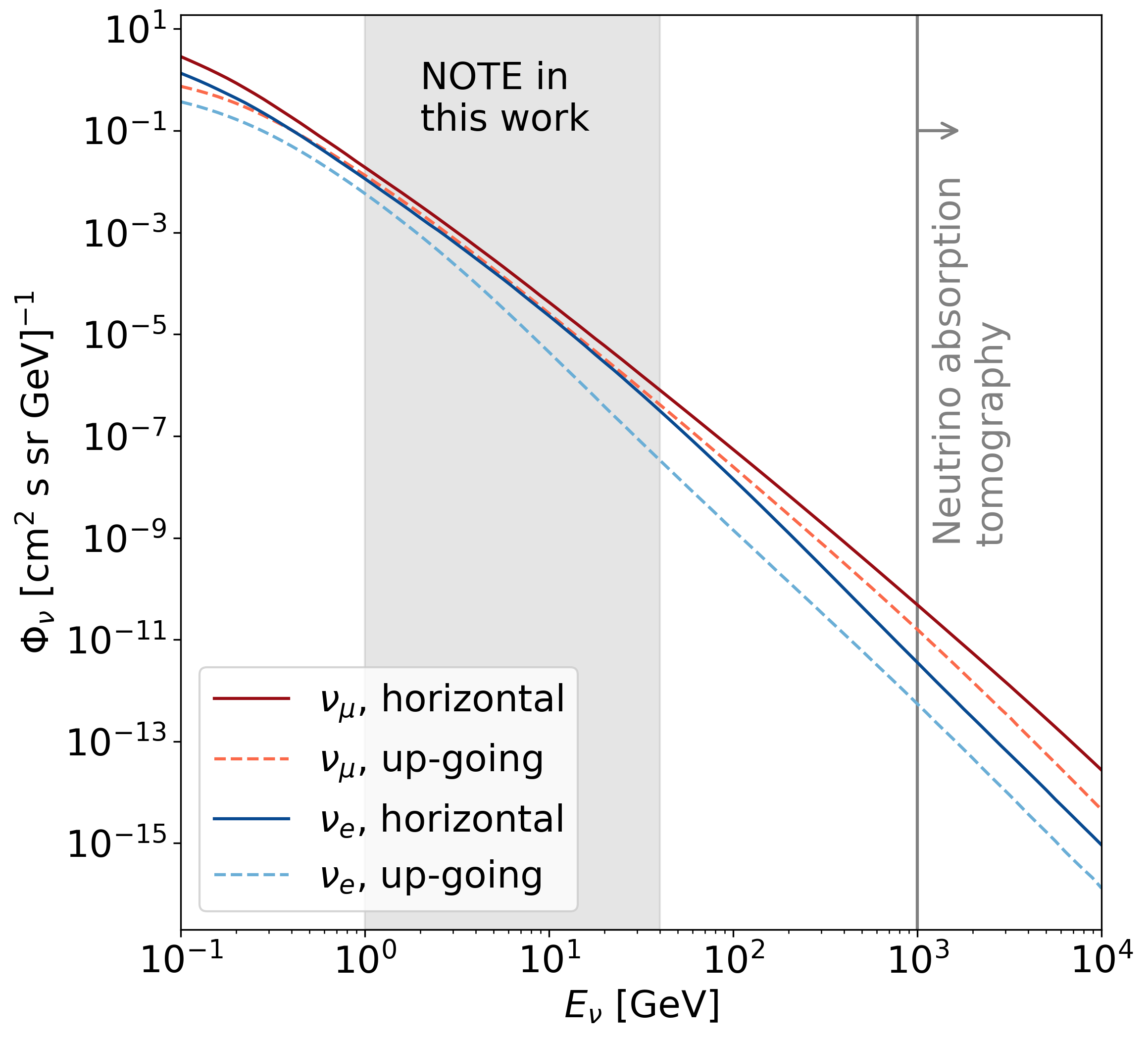}
    \caption{Flux of atmospheric electron neutrinos (blue curves) and muon neutrinos (red curves) from~\citet{honda2015atmospheric}, averaged over azimuthal angle. Horizontal neutrinos (solid lines) correspond to $\cos\theta_z$ values between $-0.1$ and $0.0$, while up-going neutrinos (dashed lines) span the range from $-1.0$ to $-0.9$. The shaded region represents the neutrino energy range considered in this work for NOTE. The vertical gray line indicates the approximate energy above which neutrino absorption becomes relevant.}
    \label{fig:honda}
\end{figure}

The atmospheric neutrino flux decreases steeply with energy, as shown in Figure~\ref{fig:honda} (the $\nu_\tau$ and $\bar{\nu}_\tau$ contributions are negligible and thus not shown). This decline emphasizes the importance of low-energy neutrinos for tomography, as they provide the larger event statistics. Neutrino absorption tomography becomes relevant above $\sim$\SI{e3}{\giga\eV} (marked by a vertical line), where the atmospheric neutrino flux is notably lower than at energies relevant for NOTE (represented by the shaded band). The flux of neutrinos produced in astrophysical point sources is even lower and only comparable with the atmospheric neutrino flux near \SI{e4}{\giga\eV}~\citep{abbasi2026evidence}. Yet, the impact of Earth model perturbations on neutrino rates is larger for absorption than for oscillations, making both methods relevant for Earth tomography. While we envision a joint inversion including both approaches, in this study we focus on NOTE.

\subsection{Neutrinos traversing the Earth}
\label{subsec:nuearth}

In this section, we outline the key aspects of flavor transitions in matter (step (b) in Figure~\ref{fig:g}). When a neutrino is produced, its quantum state collapses to a flavor eigenstate denoted $|\nu_\alpha\rangle$, where $\alpha \in \{e, \mu, \tau\}$. Neutrino oscillations arise because eigenstates of propagation are not aligned with these flavor eigenstates.

In vacuum, the propagation eigenstates are the mass eigenstates $|\nu_j\rangle$ (with $j \in \{1, 2, 3\}$), which correspond to the possible outcomes of a neutrino mass measurement, each with a definite mass $m_j$. Thus, a neutrino produced in the flavor eigenstate $|\nu_\alpha\rangle$ is a superposition of the mass eigenstates:
\begin{equation}\label{eq:U}
|\nu_\alpha\rangle = \sum_{j=1}^3 U_{\alpha j}^* |\nu_j\rangle.
\end{equation}
$U$ denotes the complex unitary Pontecorvo--Maki--Nakagawa--Sakata (PMNS) matrix, which performs the change of basis between the orthonormal sets of flavor and mass eigenstates:
\begin{equation}
U =
\begin{pmatrix}
1 & 0 & 0 \\
0 & c_{23} & s_{23} \\
0 & -s_{23} & c_{23}
\end{pmatrix}
\begin{pmatrix}
c_{13} & 0 & s_{13} e^{-i\delta_{\rm CP}} \\
0 & 1 & 0 \\
- s_{13} e^{i\delta_{\rm CP}} & 0 & c_{13}
\end{pmatrix}
\begin{pmatrix}
c_{12} & s_{12} & 0 \\
- s_{12} & c_{12} & 0 \\
0 & 0 & 1
\end{pmatrix}.
\end{equation}
Unitarity reduces its $18$ real parameters to $s_{ij} = \sin\Theta_{ij}$, $c_{ij} = \cos\Theta_{ij}$, and $\delta_{\rm CP}$, after unphysical phases are absorbed into the field definitions.

If this neutrino propagates over a distance $L$, each $|\nu_j\rangle$ evolves with a phase $\Psi$ that depends on its mass $m_j$, $E_\nu$ and $L$, which we leave in symbolic form for simplicity:
\begin{equation}\label{eq:Uevol}
|\nu_\alpha(L)\rangle = \sum_{j=1}^3 U_{\alpha j}^* \left( e^{i \Psi(m_j, E_\nu, L)} |\nu_j\rangle \right).
\end{equation}
Since these phases differ for each $j$, the superposition at distance $L$ is no longer the same as at production. Expressing $|\nu_\alpha(L)\rangle$ back in terms of the flavor eigenstates $|\nu_\beta\rangle$ gives
\begin{equation}
|\nu_\alpha(L)\rangle = \sum_{\beta=e, \mu, \tau} \left( \sum_{j=1}^3 U_{\alpha j}^* e^{i \Psi(m_j, E_\nu, L)} U_{\beta j} \right) |\nu_\beta\rangle.
\end{equation}

When the neutrino is detected, its quantum state collapses to a flavor eigenstate $|\nu_\beta\rangle$ ($\beta \in \{e, \mu, \tau\}$) with probability equal to the squared modulus of the corresponding coefficient:
\begin{equation}
\text{P}_{\nu_\alpha \rightarrow \nu_\beta}(L) = \left| \sum_{j=1}^3 U_{\alpha j}^* e^{i \Psi(m_j, E_\nu, L)} U_{\beta j} \right|^2.
\end{equation}
The complex exponentials that arise lead to sinusoidal terms in the transition probabilities, giving neutrino oscillations their name.

The expression for $\Psi$ in Equation~\eqref{eq:Uevol} is obtained by solving the Schrödinger equation (which describes the temporal evolution of the neutrino, or equivalently, its evolution with distance $L$)
\begin{equation}\label{eq:schrodinger}
i \hbar \partial_t \ket{\nu (t)} = \mathcal{H}(t) \ket{\nu (t)},
\end{equation}
for each $|\nu_j\rangle$. $\mathcal{H}(t)$ is the Hamiltonian governing the system's evolution and $\hbar$ is the reduced Planck constant ($i$ in this context refers to the complex number). In vacuum, the Hamiltonian is given by
\begin{equation}
\mathcal{H}_\text{vacuum} = \frac{1}{2 E_\nu} U 
\begin{pmatrix}
0 & 0 & 0\\
0 & \Delta m_{21}^2 & 0\\
0 & 0 & \Delta m_{31}^2
\end{pmatrix}
U^\dagger,
\end{equation}
where $\Delta m_{ij}^2 = m_i^2 - m_j^2$ and the $\dagger$ denotes the Hermitian conjugate.
If, however, the neutrino travels through matter, the effective matter Hamiltonian 
\begin{equation}\label{eq:hmatter}
\mathcal{H}_\text{matter}(t) = \begin{pmatrix}
V_e(t) & 0 & 0\\
0 & 0 & 0\\
0 & 0 & 0
\end{pmatrix}
\end{equation} 
needs to be added:
\begin{equation}\label{eq:Htotal}
\mathcal{H}(t) = \mathcal{H}_\text{vacuum} + \mathcal{H}_\text{matter}(t).
\end{equation}

Since $\mathcal{H}_\text{vacuum}$ is constant, the vacuum scenario (Equation~\eqref{eq:schrodinger}) admits simple analytical solutions, which can be found in standard neutrino textbooks~\citep[see e.g.][]{giunti2007fundamentals}. The matter scenario, which applies to NOTE, requires a numerical solution to track neutrinos as they traverse regions of varying Earth properties (comprised in Equation~\eqref{eq:hmatter}). \ref{subsec:nuearth2} outlines the treatment of this case adopted in this work and specifies the parameters entering $\mathcal{H}_\text{vacuum}$.

In NOTE, $\mathcal{H}_\text{matter}$ depends on the neutrino ray through the Earth by means of the matter-induced potential
\begin{equation}\label{eq:effpotential}
V_e(t) = \pm \sqrt{2} \, G_\mathrm{F} \, n_e(t),
\end{equation}
where $n_e(t)$ is the electron density profile along the ray, $G_\mathrm{F}$ is the Fermi constant, and the sign is positive (negative) for neutrinos (antineutrinos). Neutrino oscillation probabilities thus depend on the electron density along the neutrino ray:
\begin{equation}\label{eq:n_e}
n_e = \frac{\rho\ Y_e\ \mathcal{N}_\mathrm{A}}{M_\text{u}},
\end{equation}
where $\rho$ is the matter density, $Y_e$ is the electron yield (a dimensionless quantity representing the number of electrons per nucleon), $\mathcal{N}_\mathrm{A}$ is the Avogadro number, and $M_\text{u} = 10^{-3}$~kg/mol is the molar mass constant. The electron density defined in this way is the number of electrons per unit volume.

\begin{figure}[!htb]
    \centering
    \includegraphics[width=1.0\linewidth]{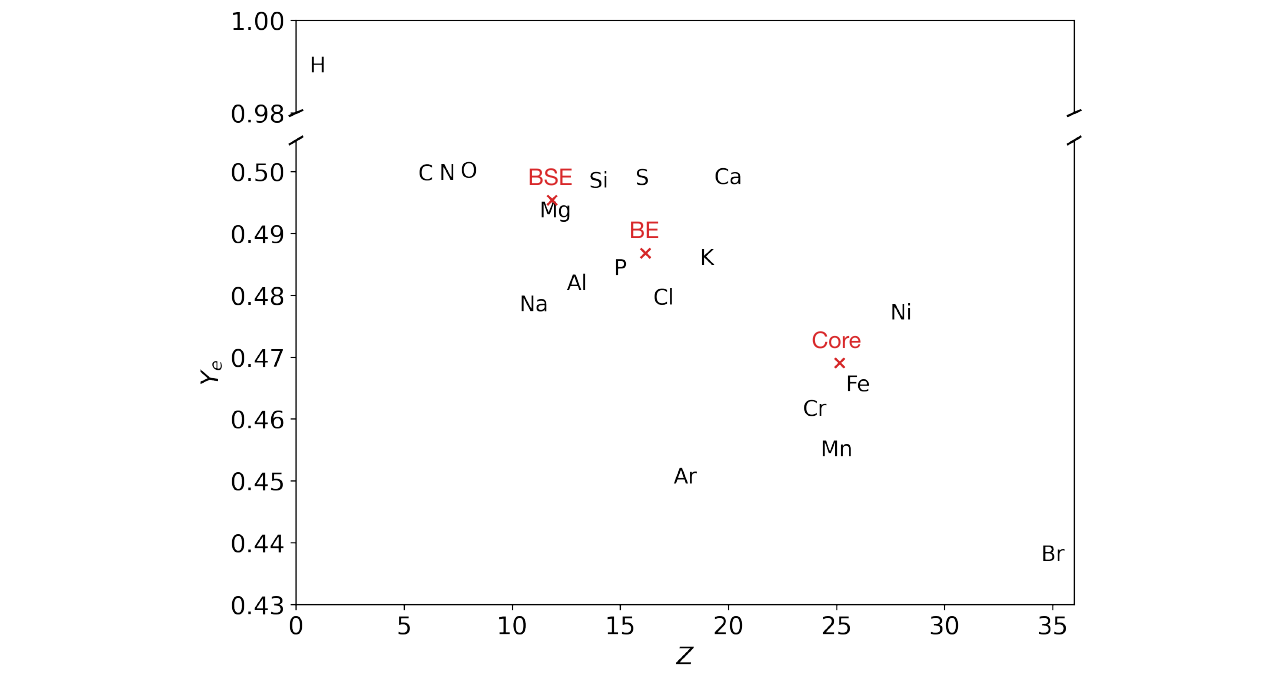}
    \caption{Electron yield $Y_e$ as a function of average atomic number $Z$ for the chemical elements considered in Equation~\eqref{eq:y_e} and for selected Earth materials. Chemical elements are indicated by their symbols (in black), while the red crosses denote the bulk Earth (BE), bulk silicate Earth (BSE), and the core. Note that, for pure chemical elements, the electron yield coincides with the ratio of atomic number $Z$ to standard atomic weight $A$. The average atomic number of the chemical elements is obtained by considering all possible isotopes~\citep{tuli1995nuclear}, while that of the composites is calculated using the element fractions from~\citet{mcdonough2010compositional}.}
    \label{fig:zoa}
\end{figure}

For our nominal (or reference) Earth model, we adopt the density profile from PREM~\citep{dziewonski1981preliminary} and compute the electron yield at each position $\bar{x}$ as:
\begin{equation}\label{eq:y_e}
Y_e(\bar{x}) = \sum_{i} w_i(\bar{x}) \bigg(\frac{Z}{A}\bigg)_i,
\end{equation}
where the sum is taken over all the elements present at $\bar{x}$. The time dependence in Equation~\eqref{eq:effpotential} is recovered using that neutrinos travel almost at the speed of light $c=|\bar{x}-\bar{x}_\text{entry}|/t$, where $\bar{x}_\text{entry}$ is the position where the neutrino enters the Earth. $w_i$ is the weight fraction of element $i$, with $Z$ and $A$ its atomic number and standard atomic weight, respectively. The composition is taken from~\citet{mcdonough2010compositional} if $\bar{x}$ is in the mantle or core, from~\citet{wedepohl1995composition} for crustal materials, and from~\citet{quinby1983distribution} for oceanic materials (\ref{sec:tables} provides a table summarizing the resulting values of $Y_e$). 

$Y_e$ reflects chemical composition, with values $\sim$0.47 in the core and $\sim$0.50 in the mantle and crust. Figure~\ref{fig:zoa} shows the $Y_e$ values for the chemical elements considered in this work (abundances above $0.01\%$) and for selected Earth materials~\citep{mcdonough2010compositional}. An overall decreasing trend with $Z$ is observed, with hydrogen standing out due to its value close to unity. This sensitivity of neutrino oscillations to composition through the electron yield further motivates the use of neutrinos to probe the Earth's interior. Equation~\eqref{eq:n_e} implies that NOTE is sensitive to the combined effect of matter density $\rho$ and composition through $Y_e$, which together define the model parameters $\mathbf{m}$. If $\rho$ were independently and sufficiently well constrained, the electron density $n_e$ could be directly related to $Y_e$. However, the level of precision required on $\rho$ for this interpretation to be valid remains to be quantified.

\subsection{Neutrino detection}
\label{subsec:nudet}

Once neutrinos reach the vicinity of the detector, they have a small probability of interacting (step (c) in Figure~\ref{fig:g}) and generating a detectable signature (step (d) in Figure~\ref{fig:g}).

In the energy range of interest for NOTE, neutrinos interact primarily by scattering off nuclei in the medium~\citep{formaggio2012ev} (step (c) in Figure~\ref{fig:g}). These interactions are classified as neutral-current (NC) or charged-current (CC), depending on whether a $Z^0$ or $W^\pm$ weak-force boson is exchanged. In NC interactions, the neutrino (antineutrino) remains in the final state and escapes the instrumented detector volume, while the scattered nucleus can produce a detectable signal. NC interactions are flavor-independent (and thus not sensitive to oscillations), giving two possible interaction channels, denoted as $\nu \, \mathrm{NC}$ and ${\bar{\nu}} \, \mathrm{NC}$. In CC interactions, the incoming neutrino (antineutrino) $\nu_\alpha$ ($\bar{\nu}_\alpha$), with $\alpha \in \{e, \mu, \tau\}$, produces a negatively (positively) charged particle $\alpha^-$ ($\alpha^+$). Both the scattered nucleus and the outgoing charged particle can produce detectable signals, resulting in six possible interaction channels: $\nu_\alpha \, \mathrm{CC}$ and $\bar{\nu}_\alpha \, \mathrm{CC}$, for $\alpha \in \{e, \mu, \tau\}$. The probability for any of these eight interaction channels to occur depends on the detection medium and scales linearly with the number of nuclear targets, i.e., the detector mass.

\begin{figure}[!htb]
    \centering
    \includegraphics[width=1\linewidth]{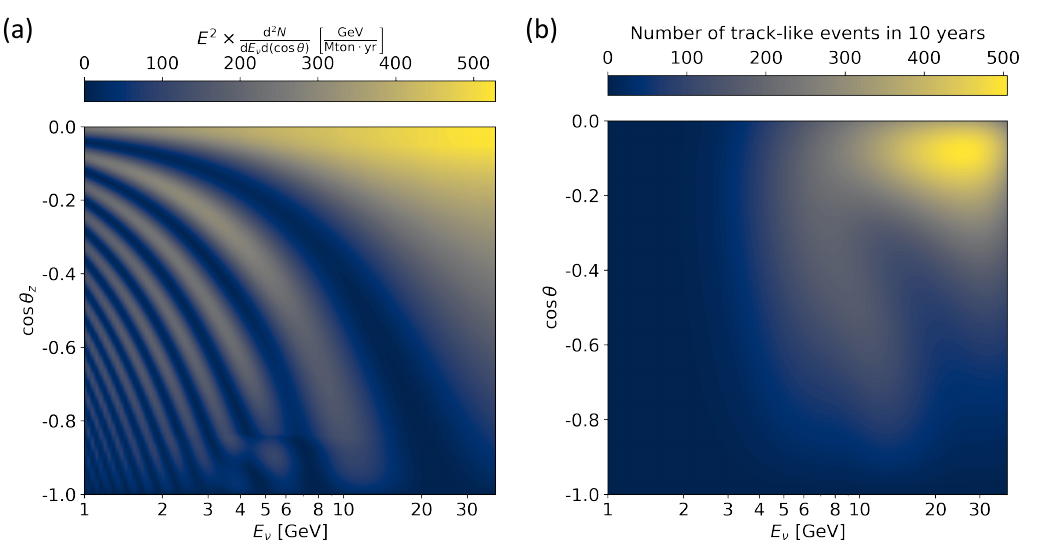}
    \caption{(a) Simulated rate of interacting neutrinos in the $(\nu_\mu + \bar{\nu}_\mu) \, \mathrm{CC}$ channels (corresponding to 1~Mton-year of exposure of a water-based detector), and multiplied by $E_\nu^2$ to visually compensate for the rapid decrease of the neutrino flux with energy. (b) Expected number of detected $\nu_\mu$-like events in a KM3NeT/ORCA-like detector with $10$~years of measurements.}
    \label{fig:IntsDets}
\end{figure}

Combining steps (a)–(c) in Figure~\ref{fig:g} yields the expected number of interacting events in a given detector (see~\ref{subsec:nuint2} for the implementation in this work). Figure~\ref{fig:IntsDets}a shows the expected distribution of $(\nu_\mu + \bar{\nu}_\mu) \, \mathrm{CC}$ events for an exposure of 1~Mton-year, defined as the product of detector mass and data-acquisition period. This quantity characterizes the event statistics and depends only on the detector material through the interaction probabilities of the different channels (a water detection medium is assumed in Fig.~\ref{fig:IntsDets}a). Comparing these distributions for different Earth models $\mathbf{m}$ helps identify those interaction channels and regions in $(E_\nu, \cos\theta_z)$ space that carry the strongest inherent discriminating signal.

In practice, observations $\mathbf{d}$ are affected by the limited efficiency and detection resolution of real detectors, where efficiency refers to the fraction of neutrino interactions that are successfully detected and resolution refers to the finite precision with which neutrino energy, direction, and flavor are inferred (i.e., reconstructed) from the measured signals.

The dominant detectable signatures in neutrino experiments arise from $\nu_e$/$\bar{\nu}_e$ and $\nu_\mu$/$\bar{\nu}_\mu \, \mathrm{CC}$ interactions. Electrons/positrons (the electron’s antiparticle) emerging from $\nu_e/\bar{\nu}_e \, \mathrm{CC}$ interactions generate event topologies (or signatures) that are qualitatively different from those induced by muons/anti-muons generated in $\nu_\mu/\bar{\nu}_\mu \, \mathrm{CC}$ interactions. Furthermore, most next-generation detectors are not designed to distinguish between neutrinos and antineutrinos on an event-by-event basis. Therefore, in a first simplification, neutrino events in this work are classified into $\nu_\mu$-like and $\nu_e$-like topologies.

Reconstruction algorithms in neutrino detectors exploit the spatial and temporal patterns of signals recorded by sensors to infer the event topology ($\nu_\mu$-like or $\nu_e$-like), as well as the neutrino energy and arrival direction (step (d) in Figure~\ref{fig:g}). Their performance is limited by detector properties such as the number, density, and efficiency of detector elements, which determine the instrument's detection efficiency, its energy and angular resolutions, and its accuracy of topology classification. Combining steps (a)–(d) in Figure~\ref{fig:g} yields the synthetic data $\mathbf{u}$, corresponding to the expected number of detected events in a given detector (see~\ref{subsec:nudet2} for the implementation in this work). Figure~\ref{fig:IntsDets}b shows the distribution of $\nu_\mu$-like events after $10$~years of measurements with the KM3NeT/ORCA water Cherenkov detector, a \mbox{3D} array of photosensors currently being deployed deep in the Mediterranean Sea, which will cover a total instrumented volume of approximately \SI{6.7e6}{\cubic\metre} once completed. Together with the corresponding distribution of $\nu_e$-like events in $(E_\nu, \cos\theta_z)$, these distributions define the synthetic data $\mathbf{u}$ for KM3NeT/ORCA. Owing to the finite energy and angular resolution of the detector, the oscillation patterns in the interacting-event distribution are smeared out, reducing the contrast of the underlying features.

\section{Methods}
\label{sec:methods}

\subsection{Forward model setup}
\label{subsec:earthprobe}

For the present work, we employ the EarthProbe framework~\citep{earthprobe}, which implements the full forward modeling of neutrino observables described in Section~\ref{sec:forward}. Within this framework, several input parameters must be provided. 

First, EarthProbe requires an electron density profile, $n_e$ (see Equations~\eqref{eq:hmatter}–\eqref{eq:y_e}). Here, it allows for user-defined $\rho$ and $Y_e$ profiles. In this work, they are discretized into $64$ shells. Given the early stage of neutrino oscillation tomography of the Earth (NOTE), we adopt this relatively coarse parameterization of the $n_e$ profile as a first step, although it can be refined in future applications. The nominal Earth model (defined in Section~\ref{subsec:nuearth}) is taken to generate the ``true'' data $\mathbf{d}$, while synthetic data $\mathbf{u}$ are produced using alternative Earth models. The alternative Earth models considered here differ from the nominal one only by a specified percentage in $n_e$ within one of the $64$ shells. We verified that the sensitivity scales linearly with the perturbation amplitude (see~\ref{subsec:lin}), and therefore, without loss of generality, we present results for a $+1\%$ modification of $n_e$.

Moreover, EarthProbe allows for different detector configurations. Each neutrino detector yields corresponding synthetic data $\mathbf{u}$, reflecting differences in detector response and reconstruction performance (steps (c) and (d) in Figure~\ref{fig:g}). The reconstruction of neutrino properties in EarthProbe follows the paradigm of water Cherenkov detectors such as KM3NeT/ORCA, as it assumes a water detection medium. EarthProbe models the detector response using a set of analytical parameterizations. This approach is sufficiently flexible to emulate technologies beyond water Cherenkov detectors. 

Regarding the state-of-the-art detectors considered, we use the benchmark parameterizations of KM3NeT/ORCA, Hyper-Kamiokande (HK)~\citep{abe2018hyper}, and DUNE~\citep{abi2020volume}, as provided in~\citet{maderer2023unveiling}. Since the parameterizations for KM3NeT/ORCA may evolve as its development progresses, and because HK and DUNE rely on a different detection technology than KM3NeT/ORCA, we refer to these configurations as KM3NeT/ORCA-like, HK-like, and DUNE-like, respectively.

To assess the intrinsic sensitivity of NOTE to perturbations in the Earth model, we also consider an idealized hypothetical detector with near-perfect reconstruction capabilities, referred to hereafter as perfect detector, since its performance is taken to reach the fundamental limits imposed by the underlying detection physics. The corresponding assumptions on efficiency, energy and angular resolution are detailed in~\ref{subsec:nudet2}. We further assume that this detector can perfectly identify all eight interaction channels introduced in Section~\ref{subsec:nudet}. Since such channel-identification performance lies beyond the reach of currently funded technologies, we also consider a more realistic configuration in which only $\nu_\mu$-like and $\nu_e$-like topologies can be distinguished, while retaining the same assumptions on efficiency and resolution.

Finally, we specify the energy range and binning scheme used for $\mathbf{u}$ and $\mathbf{d}$. For all detector configurations, we use $100$ logarithmically spaced energy bins between \SI{1}{\giga\eV} and \SI{40}{\giga\eV}, and 800 bins in $\cos\theta_z$ spanning from $-1$ to $0$. Although the sensitivity of NOTE is concentrated roughly between \SI{1}{\giga\eV} and \SI{11}{\giga\eV}, we extend the energy range to \SI{40}{\giga\eV}. This choice is motivated by the limited energy and angular resolution of detectors such as KM3NeT/ORCA, which shifts sensitivity toward higher energies. For HK and DUNE, the contribution to the total sensitivity above \SI{11}{\giga\eV} is below $0.1\%$. Thus, including this extended range has no significant impact on the results, while ensuring consistency across detector models. The contribution from energies below \SI{1}{\giga\eV} is negligible for all detectors. Given our assumption of a $20$~year data-acquisition period, we adopt this fine binning in both energy and $\cos\theta_z$, which still results in small statistical fluctuations in each bin.

\subsection{Toward the inversion of neutrino data}
\label{subsec:inv}

The forward modeling of NOTE can be written in general form~\citep{virieux2010overview} as
\begin{equation}\label{eq:g}
\mathbf{u} = \mathcal{G}(\mathbf{m}).
\end{equation}
Here, $\mathbf{m}$ denotes the set of Earth model parameters, representing the electron density profile (Section~\ref{subsec:nuearth}). In the present work, $\mathbf{m}$ is a vector with $64$ entries (Section~\ref{subsec:earthprobe}). For a given neutrino detector, the synthetic data $\mathbf{u}$ consists of event counts binned in energy and incidence angle, and further separated by event topology. The binning scheme adopted in this work is specified in Section~\ref{subsec:earthprobe}.
Formulating the inverse problem requires knowledge of the inverse of $\mathcal{G}$, which is generally not accessible.
 
One way to proceed is to linearize the forward problem around the initial Earth model $\mathbf{m}_\mathrm{init}$, which is taken to be the nominal Earth model defined in Section~\ref{subsec:nuearth}, yielding
\begin{equation}\label{eq:lin}
    \mathcal{G} (\mathbf{m}_\mathrm{init} + \Delta \mathbf{m}) \simeq \mathcal{G} (\mathbf{m}_\mathrm{init}) + \left( \nabla_\mathbf{m} \left. \mathcal{G} (\mathbf{m}) \right|_{\mathbf{m}=\mathbf{m}_\mathrm{init}} \right) \cdot \Delta \mathbf{m},
\end{equation}
where $\nabla_\mathbf{m} \mathcal{G} (\mathbf{m}) $ are the Fréchet derivatives
\begin{equation}
    \left[ \nabla_\mathbf{m} \mathcal{G} (\mathbf{m}) \right]_{\alpha i} = \left. \frac{\partial u_\alpha}{\partial m_i} \right|_{\mathbf{m}=\mathbf{m}_\mathrm{init}}.
\end{equation}
This matrix encodes the linear response of the observables to perturbations in the model parameters and therefore constitutes the central ingredient for inversion based on neutrino data. Performing an inversion with neutrinos thus requires the construction of $\nabla_\mathbf{m} \mathcal{G}(\mathbf{m})$, a subset of which we present in Section~\ref{subsec:kernels}.

In this work, $\mathbf{m}$ is a \mbox{1D} parameterization of the Earth. In seismology, such a parameterization is typically used for global or regional \mbox{1D} inversion, since seismic waves sample extended volumetric regions between source and receiver. In contrast, neutrino rays are straight lines, intersecting a spherically symmetric shell under study at up to two points (one in the tangential case). Consequently, in NOTE, the Fréchet derivatives $\left. \partial u_\alpha / \partial m_i \right|_{\mathbf{m}=\mathbf{m}_\mathrm{init}}$ can be interpreted as encoding sensitivity to localized 3D voxels along the ray. A \mbox{3D} inversion would require resolving the degeneracy between the two voxels located at the same radial distance from the Earth’s center along a given neutrino ray. This step is beyond the scope of this work, but could be addressed by introducing a hemispherical partition of the Earth relative to the detector position. We present Fréchet derivatives that do not account for this distinction.

\subsection{Statistical approach to NOTE}
\label{subsec:chi2}

Fréchet derivatives constitute the tool for performing inversion. Inspecting them can offer qualitative insight into which regions in $(E_\nu, \cos\theta_z)$ carry stronger predictive power, by identifying energies and incidence angles where $\left.\partial u_\alpha / \partial m_i \right|_{\mathbf{m}=\mathbf{m}_\mathrm{i}}$ is large. However, this interpretation remains only indicative. A large variation in $\mathbf{u}$ in a given bin does not necessarily imply a significant contribution if the corresponding event count has large uncertainty. A quantitative assessment requires accounting for the statistical uncertainty associated with the bin populations.

Consequently, as a measure of the sensitivity of the idealized and realistic next-generation detectors described in Section~\ref{subsec:earthprobe} to distinguish two different electron density scenarios, we report in Sections~\ref{subsec:sensitivity1} and~\ref{subsec:sensitivity2} the total $\Delta \chi^2$, computed as
\begin{equation}\label{eq:Delta_chi2}
    \Delta \chi^2 (\mathbf{m}_\mathrm{init}, \mathbf{m}_\mathrm{mod, i}) = \sum_{j=1}^{n_E} \sum_{k=1}^{n_\theta}\frac{\left( N_{j,k}^\mathrm{init} - N_{j,k}^\mathrm{mod,i} \right)^2}{N_{j,k}^\mathrm{init}},
\end{equation}
where the summation is over the $n_E$ energy and $n_\theta$ incidence angle bins. $\mathbf{m}_\mathrm{mod, i} = \mathbf{m}_\mathrm{init} + \Delta \mathbf{m}$ refers to the alternative Earth model, that differs from the nominal one by a modification of $n_e$ in the $i$th shell. $N_{j,k}^\mathrm{init}$ and $N_{j,k}^\mathrm{mod, i}$ denote the numbers of detected neutrino events for the initial and modified Earth models, respectively, and follow Poisson statistics. The binning described in Section~\ref{subsec:earthprobe} is sufficiently fine to approximate the limit of infinitely small bins. The values $\Delta \chi^2 (\mathbf{m}_\mathrm{init}, \mathbf{m}_\mathrm{mod, i})$ should be interpreted as the statistical significance with which the nominal model would be rejected if the event counts $N_{j,k}^\mathrm{mod, i}$ in Equation~\eqref{eq:Delta_chi2} were observed.

\subsection{Physical origin of the sensitivity of NOTE}
\label{subsec:msw}

The sensitivity of NOTE depends on how strongly variations in the Earth model affect the neutrino rates. The model parameters $\mathbf{m}$ directly influence the neutrino oscillation probabilities, which modulate the flavor composition of the atmospheric neutrino flux traversing the Earth and thereby translate into changes in the observed neutrino rates across flavors.

In the simplified case of a uniform Earth with constant electron density $n_e^0$, oscillation probabilities are proportional to $\sin^2{(c(n_e^0) \cdot L/E_\nu)}$, where $L$ is the propagation distance and $c(n_e^0)$ is a constant that depends on the electron density. For specific combinations of $L$ and $E_\nu$, this term varies strongly between Earth models, leading to large changes in oscillation probabilities, thus neutrino rates, and therefore increased sensitivity. These conditions correspond to resonant oscillations, known as MSW resonances (for Mikheyev, Smirnov, and Wolfenstein). In the Earth, such resonances occur for neutrinos with $\sim$\SI{3}{\giga\eV} traversing the outer core and with $\sim$\SI{7}{\giga\eV} traversing the mantle with $\sim$\SI{7}{\giga\eV}. 

Neutrino flavor oscillations within the Earth can undergo another form of resonant enhancement, known as parametric resonance, arising from density transitions between different terrestrial layers~\citep{chizhov1998oscillation, akhmedov1999parametric}. In this case, effects accumulated in successive regions can combine in a constructive manner, leading to a significant modification of the final oscillation probability and increased sensitivity.

\begin{figure}[!htb]
    \centering
    \includegraphics[width=1\linewidth]{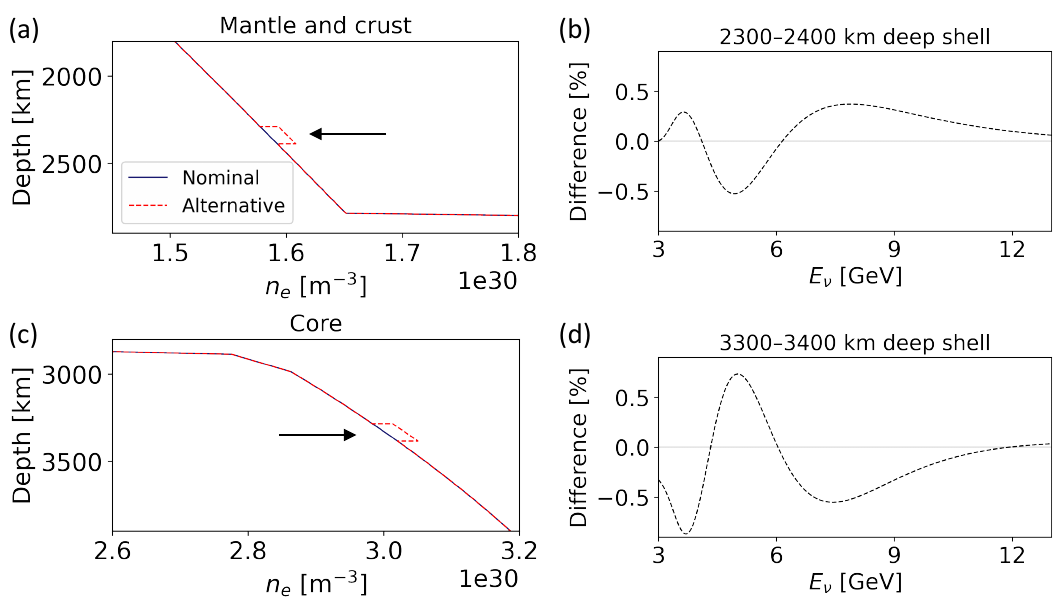}
    \caption{Electron density profile of the (a) mantle and crust and of the (c) core, (see Section~\ref{subsec:nuearth}, solid blue line). In (a), the red dashed curve shows the alternative Earth model in which the electron density is increased only between $\sim$\SI{2300}{\kilo\m} and $\sim$\SI{2400}{\kilo\m}. In (c), the same modification is applied between $\sim$\SI{3300}{\kilo\m} and $\sim$\SI{3400}{\kilo\m}. Panels (b) and (d) show the corresponding differences (in $\%$) in P$_{\nu_e \rightarrow \nu_\mu}$ between the nominal and modified Earth models from (a) and (c), respectively. In each case, only neutrinos traversing the modified shell tangentially are considered.}
    \label{fig:n_e}
\end{figure}

The interplay of both resonance mechanisms gives rise to energy ranges in which neutrinos traversing specific regions of the Earth undergo enhanced modifications of their oscillation probabilities. Figure~\ref{fig:n_e} illustrates the effect of increasing by $1\%$ the electron density in a shell located either in the mantle (\ref{fig:n_e}a-b) or in the core (\ref{fig:n_e}c-d) on the oscillation probability P$_{\nu_e \rightarrow \nu_\mu}$, for neutrinos propagating tangentially through each shell. In the core, a variation in electron density could reflect differences in light-element alloy composition, such as FeH versus FeX (with X$\neq$H). The deviation between the nominal and modified Earth models is generally larger (in absolute value) when the perturbation is located in the core.

\section{Sensitivity of NOTE to variations in the radial electron density}
\label{sec:results}

\subsection{Fréchet derivatives}
\label{subsec:kernels}

\begin{figure}[!htb]
    \centering
    \includegraphics[width=1\linewidth]{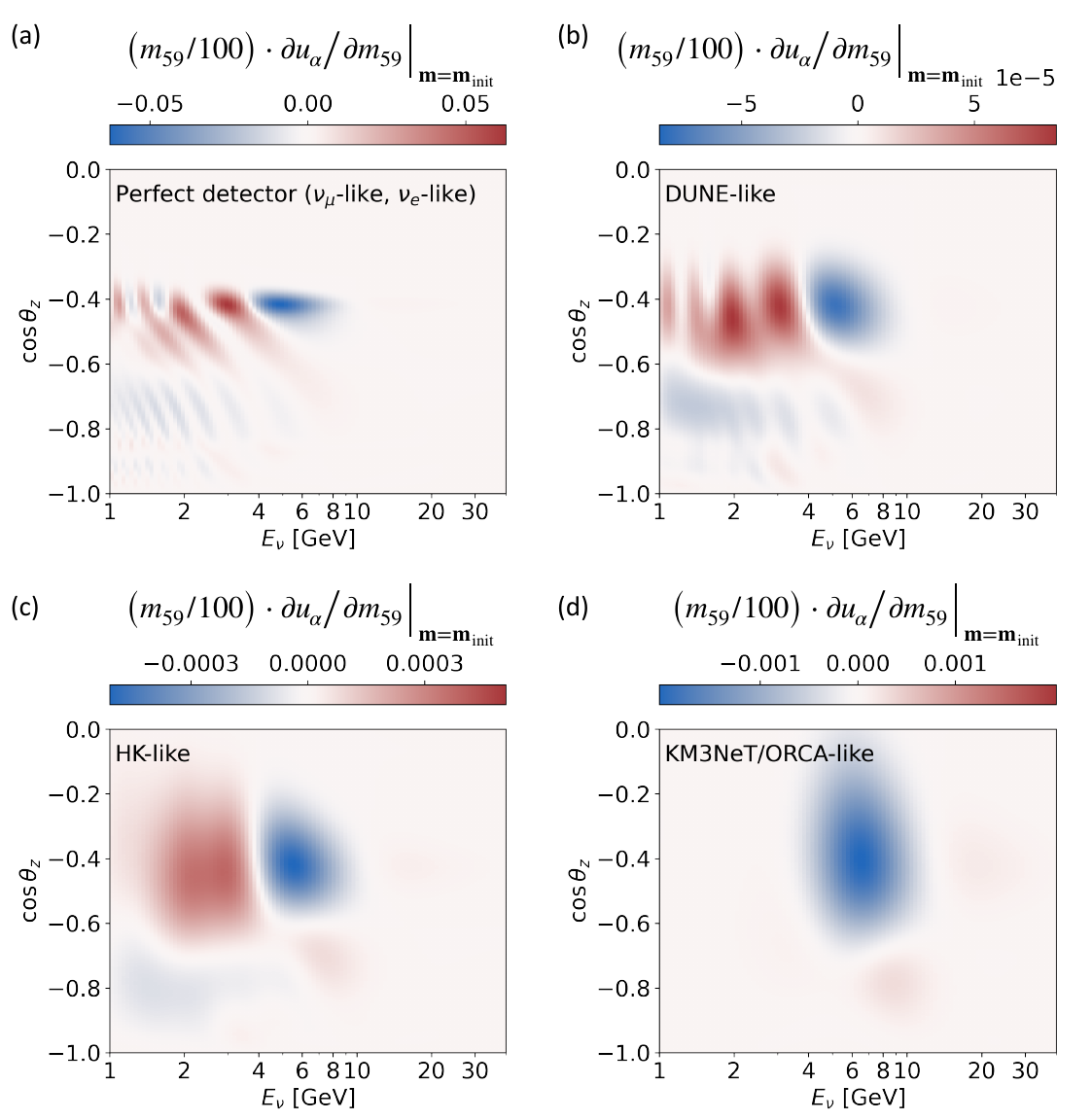}
    \caption{Fréchet derivatives $\left. \partial u_\alpha \big/ \partial m_i \right|_{\mathbf{m}=\mathbf{m}_\mathrm{init}}$ in the inversion for electron density, with the perturbation applied as a percentage variation and normalized to $m_{59}$. The results are shown for a $\sim$\SI{100}{\kilo\meter}-thick shell within the MTZ, spanning depths from \SI{500}{\kilo\meter} to \SI{600}{\kilo\meter} and parameterized by $m_{59}$. Results are shown for $\nu_\mu$-like events in the perfect (a), DUNE-like (b), HK-like (c), and KM3NeT/ORCA-like (d) detectors. The derivatives correspond to a $+1\%$ perturbation in electron density.}
    \label{fig:kernels}
\end{figure}

In this section, we present a subset of the Fréchet derivatives introduced in Section~\ref{subsec:inv} that form the basis of a future linearized inversion. Figure~\ref{fig:kernels} shows these derivatives with respect to a density change in a $\sim$\SI{100}{\kilo\meter}-thick shell within the mantle transition zone (MTZ). The corresponding Fréchet derivatives are shown for the detector configurations used in this work, across all energy and incidence-angle bins, and restricted to $\nu_\mu$-like events. In NOTE, Fréchet derivatives depend on energy, incidence angle, and flavor (encoded through event topologies). Furthermore, as required for the linearized inversion, they are linear in the perturbation amplitude (see~\ref{subsec:lin}).

For the perfect detector, $\left. \partial u_\alpha \big/ \partial m_i \right|_{\mathbf{m}=\mathbf{m}_\mathrm{init}}$ vanishes above \SI{500}{\kilo\meter}, since neutrinos that do not cross the \SI{500}{\kilo\meter}–\SI{600}{\kilo\meter} shell are unaffected. The largest values occur for neutrinos crossing this shell tangentially (at $\cos\theta_z \simeq -0.4$, or equivalently $\Delta \simeq 48^\circ$), which maximizes their path length through the perturbed region. For more vertical rays, the shell is crossed twice but over shorter path lengths, reducing the effect. In the DUNE-like detector, finite energy and angular resolutions induce event migration between bins, in particular into angular bins above the perturbed shell, and leading to a smearing of the pattern observed in Figure~\ref{fig:kernels}b. This effect becomes progressively stronger for the HK-like and KM3NeT/ORCA-like configurations, whose resolutions are comparatively poorer. 

Additionally, the range of $\left. \partial u_\alpha \big/ \partial m_i \right|_{\mathbf{m}=\mathbf{m}_\mathrm{init}}$ values varies across detectors, increasing with detector size. Consequently, for the KM3NeT/ORCA-like configuration, although the fine structure of the pattern from Figure~\ref{fig:kernels}a is degraded by limited resolution, the residual signal remains sufficiently strong to make such a detector potentially relevant for NOTE.

\subsection{Sensitivity dependence on energy and incidence angle}
\label{subsec:sensitivity1}

\begin{figure}[!htb]
    \centering
    \includegraphics[width=1\linewidth]{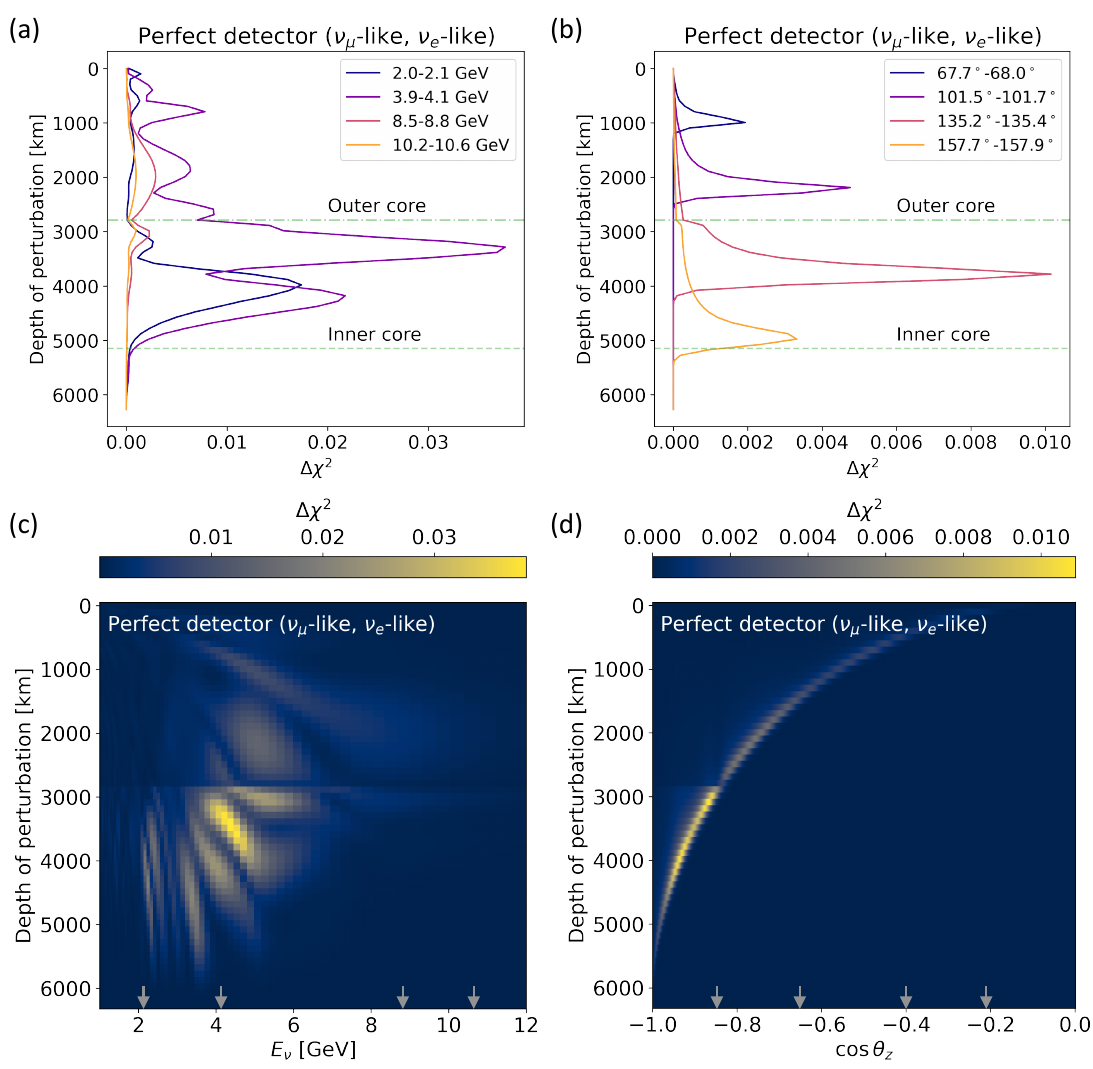}
    \caption{Sensitivity curves, as defined in Equation~\eqref{eq:Delta_chi2}, for $+1\%$ perturbations in electron density within \SI{100}{\kilo\meter}-thick shells at different depths. Panels (a) and (c) show the energy dependence for selected energies and for all energies, respectively, while panels (b) and (d) show the corresponding incidence-angle dependence for selected angles and for all angles. The gray arrows in panels (c) and (d) indicate the selected energies shown in (a) and the selected incidence angles shown in (b), respectively. For guidance, the boundaries of the outer core and inner core are also shown in panels (a) and (b).}
    \label{fig:sensitivityEct}
\end{figure}

In this section and the next, we assess the sensitivity of NOTE to perturbations in individual shells, as described in Section~\ref{subsec:chi2}. To assess the energy and angular dependence of NOTE sensitivity in a perfect detector, we use Equation~\eqref{eq:Delta_chi2}, first summing over all $n_\theta=800$ incidence-angle bins considered in this work while retaining a single energy bin ($n_E=1$, Figures~\ref{fig:sensitivityEct}a and~c) and second summing over all $n_E=100$ energy bins while retaining a single incidence-angle bin ($n_\theta=1$, Figures~\ref{fig:sensitivityEct}b and d). Each depth on the y-axis corresponds to the comparison between neutrino rates computed with the nominal Earth model and an alternative model $\mathbf{m}_\mathrm{mod,i}$, while the x-axis shows the computed $\Delta\chi^2$ values.

We find that the sensitivity for neutrinos with energies around \SI{4}{\giga\eV} is highest to perturbations in the outer core (Figures~\ref{fig:sensitivityEct}a and c). This behavior is driven by the MSW resonance experienced by low-energy neutrinos traversing it. The MSW resonance around \SI{7}{\giga\eV} in the mantle enhances the sensitivity to mantle perturbations for neutrinos of higher energy, while NOTE sensitivity becomes negligible for neutrino energies above $\sim$\SI{11}{\giga\eV}.

Neutrinos in a narrow incidence-angle bin are most sensitive to perturbations of the shell they cross tangentially (Figures~\ref{fig:sensitivityEct}b and d). If the perturbed shell is deeper and not traversed, the contribution vanishes, and if it is shallower and crossed twice over short paths, $\Delta\chi^2$ is small but non-zero. As a result, the sensitivity curves peak at the depth corresponding to tangential crossing. The peak heights reflect the combined contribution of all energies.

\subsection{Total sensitivity}
\label{subsec:sensitivity2}

\begin{figure}[!htb]
    \centering
    \includegraphics[width=1\linewidth]{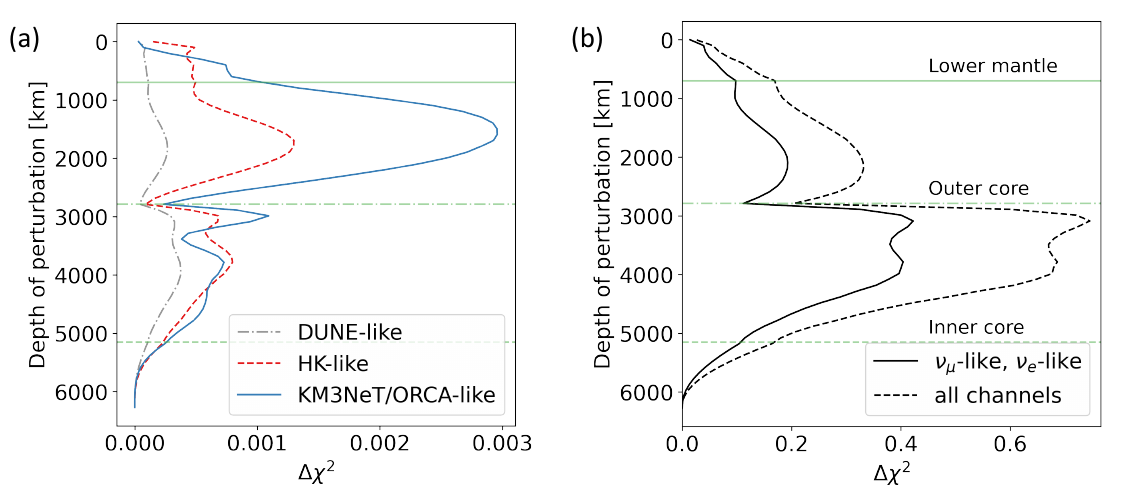}
    \caption{Total sensitivity curves, as defined in Equation~\eqref{eq:Delta_chi2}, for $+1\%$ electron density perturbations in \SI{100}{\kilo\meter}-thick shells at different depths. The results for the next-generation detectors DUNE-like, HK-like, and KM3NeT/ORCA-like are shown in (a). A data-acquisition period of $20$~years is assumed for all cases, corresponding to exposures of $0.8$, $8$, and $160$~Mton-years, respectively, reflecting the different detector sizes ($0.04$, $0.4$, and $8$~Mton, respectively). The intrinsic sensitivity is illustrated in (b) using the perfect detector capable of distinguishing all eight neutrino interaction channels (dashed curve). A more realistic scenario is also shown (solid curve), resorting to the perfect detector distinguishing only two event topologies. In these two cases an exposure of 200~Mton-years is assumed. For guidance, the boundaries of the lower mantle, outer core, and inner core are also shown.}
    \label{fig:1dsensitivity}
\end{figure}

Finally, we examine the total sensitivity to $+1\%$ electron density perturbations in \SI{100}{\kilo\meter}-thick shells by summing over all $n_E=100$ energy and $n_\theta=800$ incidence-angle bins   (Figure~\ref{fig:1dsensitivity}). These sensitivity curves demonstrate the impact of including NOTE in an inversion framework as well as the performance of next-generation detectors (Figure~\ref{fig:1dsensitivity}a) relative to a perfect detector (Figure~\ref{fig:1dsensitivity}b), thereby indicating where improvements in experimental performance would most effectively enhance sensitivity.

Each sensitivity curve in Figure~\ref{fig:1dsensitivity}a can be loosely (as systematic uncertainties are not included) interpreted as a lower bound on sensitivity for the corresponding next-generation detector due to the following. A $1\%$ increase in electron density within a shell represents a local perturbation that is not physically self-consistent, as it violates hydrostatic equilibrium (HE). In a realistic scenario, such a change would require correlated adjustments in the surrounding structure to preserve the Earth’s total mass, moment of inertia, and HE. These coupled modifications would lead to a larger deviation from the nominal model, resulting in stronger changes in neutrino event rates and thus enhanced sensitivity.

In contrast, the dashed curve in Figure~\ref{fig:1dsensitivity}b, corresponding to the perfect detector capable of distinguishing all eight neutrino interaction channels, represents an upper bound and can be interpreted as the intrinsic sensitivity of NOTE to these perturbations. The solid curve corresponds to a more realistic perfect detector distinguishing only two event topologies. In both cases, the sensitivity to increases in electron density is highest in the Earth’s outer core. It decreases sharply toward the Earth’s center due to the reduced number of neutrinos probing progressively smaller shells and the shorter path lengths through them. Sensitivity in the mantle is significantly lower, mainly because lower-density shells correspond to higher resonance energies where the atmospheric neutrino flux is reduced.

The gap between the sensitivity of next-generation detectors and the idealized case reflects limitations in event reconstruction. The reduction in sensitivity is larger in the core because neutrinos traversing it experience MSW resonances near \SI{3}{\giga\eV}, compared to $\sim$\SI{7}{\giga\eV} for mantle-crossing neutrinos, and reconstruction performance degrades at lower energies for all detectors. The effect is largest for KM3NeT/ORCA, intermediate for HK, and smallest for DUNE, reflecting their respective resolutions and efficiencies~\citep{maderer2023unveiling}.

\section{Conclusions}
\label{sec:conclusions}

Research on the Earth’s internal structure has advanced greatly over the past century, yet key questions remain, such as the precise density and hydrogen content of different regions, motivating the search for new observables that complement traditional geophysical methods. One promising observable is the flux of atmospheric neutrinos traversing the Earth, whose oscillation patterns carry information about the electron density along their rays. In this paper, we take a step toward integrating neutrino oscillation measurements into a joint inversion framework with seismic data by describing the forward problem, formulating its linearization, and quantifying the sensitivity of the associated Fréchet derivatives to perturbations in the Earth’s electron density profile.

We first describe the forward modeling of neutrino oscillation tomography of the Earth (NOTE). We then demonstrate that the Fréchet derivatives required for the linearized inverse problem can be computed and that they scale linearly with the amplitude of perturbations in electron density. This is done for expected neutrino rates at the next-generation KM3NeT/ORCA detector, which is under construction and steadily accumulating data, as well as for HK and DUNE, which are scheduled to begin data acquisition in the coming years. We also consider an idealized hypothetical detector to assess the intrinsic sensitivity of the method.

To investigate the sensitivity of the Fréchet derivatives to localized perturbations, we study the response of neutrino oscillations to a $1\%$ increase in electron density within $\sim$100~km-thick shells. To approximate the intrinsic sensitivity, we consider the idealized detector representing the fundamental performance limit. While this idealized detector shows the highest sensitivity to perturbations in the core, realistic detectors exhibit overall reduced sensitivity that shifts toward the mantle. Thus, detector performance plays an important role in determining which regions of the Earth can be probed most effectively. The sensitivity decrease arises from reduced detection efficiency and finite energy and angular resolution and is more pronounced at lower energies, where MSW resonances occur for core-traversing neutrinos. Future work should identify which detector characteristics most effectively enhance sensitivity to different Earth regions, thereby guiding the design of next-generation instruments.

These results establish the basis for developing a joint inversion framework combining seismic data and neutrino oscillation tomography of the Earth, with the latter providing constraints on the Earth's electron density distribution. Such constraints on electron density could help test specific geochemical hypotheses, such as localized hydrogen enrichment or proposed heavy-element concentrations. In its most complete form, such a framework would integrate neutrino observations from multiple detectors as well as neutrino absorption tomography, providing more comprehensive and complementary constraints on the Earth’s internal structure.

\section*{Acknowledgements} 
We would like to express our sincere gratitude to the editor and the reviewers for their insightful comments, constructive feedback, and helpful suggestions. João A. B. Coelho, Nobuaki Fuji, Isabel Goos and Véronique Van Elewyck were supported by the program ``Investissement d'Avenir'', launched by the French Government and implemented by ANR (ANR‐21‐EXES‐0002, Fire-UP Program ``Crossing Cutting Edges'' and InIdex project HERMES ``Multi-messengers of the Earth and the Universe''). Nobuaki Fuji was supported by the Institut Universitaire de France. Eric Mittelstaedt and Yael Deniz were supported by the Award 2406115 from the United States National Science Foundation. 

\appendix

\section{Physical constants, variables, and expressions}
\label{sec:tables}

In this appendix, we provide tables listing the physical constants and variables used in this study.

\begin{table}[h!]
\centering
\caption{Table of physical constants used in this study and their values. The neutrino oscillation parameters ($\Theta_{ij}$, $\Delta m_{ij}^2$, and $\delta_{\rm CP}$) are taken from~\citep{esteban2025nufit}. Other constants are taken from~\citet{pdg2015}.}
\begin{tabular}{lll}
\hline
\textbf{Symbol} & \textbf{Description} & \textbf{Value} \\
\hline
$\Theta_{12}$ & ``solar'' mixing angle & $33.68^\circ$ \\
$\Theta_{13}$ & ``reactor'' mixing angle & $8.56^\circ$ \\
$\Theta_{23}$ & ``atmospheric'' mixing angle & $43.3^\circ$ \\
$\Delta m_{21}^2$ & ``solar'' mass-squared difference & \SI{7.49e-5}{\eV\squared} \\
$\Delta m_{31}^2$ & \makecell[l]{``atmospheric'' \\ mass-squared difference} & \SI{2.513e-3}{\eV\squared} \\
$\delta_\mathrm{CP}$ & CP-violating phase & $212^\circ$ \\
$M_\text{u}$ & molar mass constant & \SI{1}{\g\per\mol}\\$\mathcal{N}_\text{A}$ & Avogadro number & \SI{6.022140857e23}{\per\mol} \\
$\hbar c$ & \makecell[l]{reduced Planck constant \\ $\times$ speed of light} & \SI{1.973269788e-10}{\eV\km} \\
$G_\text{F}$ & Fermi constant & \SI{1.1663787e-5}{\giga\per\eV\squared} \\
\hline
\end{tabular}
\label{tab:constants}
\end{table}

\begin{table}[h!]
\centering
\caption{Table of variables used in this study and their units.}
\begin{tabular}{lll}
\hline
\textbf{Symbol} & \textbf{Description} & \textbf{Unit} \\
\hline
$E_\nu$ & neutrino energy & \SI{}{\giga\eV}\\
$\theta_z$ & neutrino zenith angle & dimensionless \\
$\Delta$ & epicentral distance & $^\circ$ \\
$t,T$ & travel time of the neutrino within the Earth & \SI{}{\s} \\
$L$ & path length & \SI{}{\km} \\
$n_e$ & electron density & \SI{}{\per\m\cubed} \\
$\rho$ & matter density & \SI{}{\kilo\g\per\m\cubed} \\
$Y_e$ & electron yield & dimensionless \\
$Z$ & atomic number & dimensionless \\
$A$ & standard atomic weight & dimensionless \\
$w$ & weight fraction & dimensionless \\ 
$\phi_\nu$ & flux of atmospheric neutrinos & (\SI{}{\centi\m\squared\s\steradian\giga\eV})$^{-1}$\\
Mw & moment magnitude scale & dimensionless \\
\hline
\end{tabular}
\label{tab:vars}
\end{table}

\begin{table}[h!]
\centering
\caption{Values of electron yield used in this work.}
\begin{tabular}{lll}
\hline
\textbf{Layer} & \textbf{Radial extent [km]} & \textbf{Electron yield} $Y_e$ \\
\hline
Sea & $6368.0-6371.0$ & $0.5524$ \\
Crust & $6346.6-6368.0$ & $0.4956$ \\
Mantle & $3480.0-6346.6$ & $0.4954$ \\
Core & $0-3480.0$ & $0.4691$ \\
\hline
\end{tabular}
\label{tab:Y_e}
\end{table}

\begin{table}[h!]
\centering
\caption{Expressions used in this study and their meaning.}
\begin{tabular}{ll}
\hline
\textbf{Symbol} & \textbf{Description} \\
\hline
$\nu_e$, $\bar{\nu}_e$ & electron neutrino, electron antineutrino \\
$\nu_\mu$, $\bar{\nu}_\mu$ & muon neutrino, muon antineutrino \\
$\nu_\tau$, $\bar{\nu}_\tau$ & tau neutrino, tau antineutrino \\
$\nu/\bar{\nu}$ & generic neutrino/antineutrino, if the flavor doesn't \\
& matter \\
$\nu_X$ & representative of any of the eight neutrino interaction \\
& channels \\
$\alpha$, $\beta$ & reserved for flavor $e$, $\mu$, or $\tau$ \\
$\alpha^-$/$\alpha^+$ & reserved for particle $e$, $\mu$, or $\tau$/ antiparticle of $e$, $\mu$, \\
& or $\tau$ \\
$\nu_\alpha \rightarrow \nu_\beta$ & oscillation process from $\nu_\alpha$ to $\nu_\beta$ \\
P$_{\nu_\alpha \rightarrow \nu_\beta}$ & probability of the oscillation process from $\nu_\alpha$ to $\nu_\beta$ \\
$\ket{\nu_\alpha}$ & neutrino flavor eigenstates \\
$\ket{\nu_i}$, $i\in\{1, 2, 3\}$ & neutrino mass eigenstates \\
$m_i$, $i\in\{1, 2, 3\}$ & neutrino masses \\
$\mathcal{H}$ & Hamiltonian governing the system's evolution \\
$\mathcal{H}_\mathrm{vacuum}$ & Hamiltonian in vacuum \\
$\mathcal{H}_\mathrm{matter}$ & effective matter Hamiltonian \\
$V_e$ & matter-induced potential \\
$\bar{x}$ & position in space \\
$\bar{x}_\mathrm{entry}$ & position where a neutrino enters the Earth \\
NC/CC & neutral-current/charged-current (interaction) \\
$Z^0$, $W^\pm$ & weak-force bosons \\
$\nu_\mu$/$\nu_e$-like & event topologies \\
$\mathcal{G}$ & function encoding the forward modeling \\
$\mathbf{m}$ & Earth model parameters, with elements $m_i$ \\
$\mathbf{u}$ & synthetic data, with elements $u_\alpha$ \\
$\mathbf{d}$ & real data \\
$\nabla_\mathbf{m} \mathcal{G}(\mathbf{m})$ & Fréchet derivatives \\
$\mathbf{m}_\mathrm{init}$ & nominal or initial Earth model \\
$\mathbf{m}_\mathrm{mod,i}$ & alternative Earth model where the $i$th shell was \\ 
& modified \\
$\Delta \mathbf{m}$ & $=\mathbf{m}_\mathrm{mod,i} - \mathbf{m}_\mathrm{init}$ \\
\hline
\end{tabular}
\label{tab:expr1}
\end{table}

\begin{table}[h!]
\centering
\caption{Continuation of Table~\ref{tab:expr1}.}
\begin{tabular}{ll}
\hline
\textbf{Symbol} & \textbf{Description} \\
\hline
$\Delta \chi^2$ & test statistic defined in Equation~\eqref{eq:Delta_chi2} \\
$n_E$/$n_\theta$ & number of energy bins/incidence-angle bins \\
$N_{j,k}^\mathrm{init}$/$N_{j,k}^\mathrm{mod, i}$ & number of detected neutrino events for the initial/modified \\
& Earth model \\
$\mathcal{T}$ & time-ordering operator \\
$\Delta t_k$, $t_k$ & time intervals, intermediate values \\
$N_{\nu_X}^\mathrm{int}$/$R_{\nu_X}^\mathrm{int}$ & number/rate of interacting neutrino events \\
$T_\mathrm{det}$/$M_\mathrm{det}$ & data-acquisition time/mass of a neutrino detector \\
$\sigma_{\nu_X}^\mathrm{int}$ & cross section of neutrino interaction $\nu_X$ \\
$m_\mathrm{N}$ & average mass of target nucleons in a medium \\
$\sigma (E_\nu)$ & smearing function representing energy resolution \\
$\sigma (\theta_z; E_\nu)$ & smearing function representing angular resolution \\
$\epsilon_\mathrm{det} (E_\nu)$ & efficiency of successful detection \\
$\epsilon_\mathrm{class} (E_\nu)$ & efficiency of successful classification \\
$n_e^0$ & particular constant value of electron density \\
$c(n_e^0)$ & constant expression that depends on $n_e^0$ \\
$\Psi$ & variable expression that depends on $m_j$, $E_\nu$, and $L$ \\
\hline
\end{tabular}
\label{tab:expr2}
\end{table}

\section{Details of neutrino oscillation probability calculations}
\label{app:math}

This appendix presents additional technical material on the modeling of neutrino observables, for which we employ the EarthProbe framework~\citep{earthprobe}.

\subsection{Atmospheric neutrino flux configuration}
\label{subsec:nuflux2}

This appendix provides the specific atmospheric neutrino flux configuration used in the present study.

In this work, we use the benchmark atmospheric neutrino flux computed for the Gran Sasso site (no mountain overburden), averaged over azimuth and seasonal variations, and assuming minimum solar activity~\citep{hondatables}. Gran Sasso, which hosts the underground laboratory~\citep{lngs}, lies at a latitude similar to that of the detectors considered here. Flux differences between these sites are expected to be within $\sim$1\%~\citep{honda2015atmospheric}. Averaging over seasons is appropriate, as NOTE will rely on data collected over several years. Solar activity modulates the flux of low-energy cosmic rays entering the atmosphere, thereby affecting the production of atmospheric neutrinos. This effect remains below $1\%$ for neutrino energies above \SI{3}{\giga\eV} and reaches at most $3\%$ at lower energies~\citep{honda2015atmospheric}. The overall neutrino flux has an estimated systematic uncertainty of $7\%$ below \SI{10}{\giga\eV}, and $14\%$ in the range between \SI{10}{\giga\eV} and \SI{100}{\giga\eV}.

\subsection{Implementation of neutrino oscillations}
\label{subsec:nuearth2}

This appendix specifies the parameters entering $\mathcal{H}_\mathrm{vacuum}$ and outlines the solution of the Schrödinger equation for neutrino oscillations in matter adopted in the present work.

The vacuum Hamiltonian $\mathcal{H}_\mathrm{vacuum}$ is fully determined by six parameters: the mixing angles $\Theta_{12}$, $\Theta_{13}$, and $\Theta_{23}$, the mass-squared differences $\Delta m_{21}^2$ and $\Delta m_{31}^2$ (with $\Delta m_{ij}^2 = m_i^2 - m_j^2$), and the complex phase $\delta_{\rm CP}$. We adopt the oscillation parameters from~\citet{esteban2025nufit}, based on global fits performed by the NuFIT collaboration and regularly updated with new experimental data. The first four parameters are now known to percent-level precision, whereas $\delta_{\rm CP}$ is still poorly constrained and the sign of $\Delta m_{31}^2$ remains undetermined. Consequently, the neutrino mass ordering, whether $m_1 < m_2 \ll m_3$ (normal ordering) or $m_3 \ll m_1 < m_2$ (inverted ordering), is still unknown. This unknown physical property affects neutrino oscillation probabilities and therefore impacts this study. Throughout our analysis, we assume the normal mass ordering, which is currently slightly favored by global data~\citep{esteban2025nufit}. NOTE requires long data-acquisition periods. By the time sufficient data have been accumulated to probe the deep Earth with neutrinos, all six oscillation parameters are expected to be known with high precision.

To solve the Schrödinger equation in matter for a given neutrino ray, defined by Equations~\eqref{eq:schrodinger}-\eqref{eq:y_e}, EarthProbe builds a sequence of the Earth matter segments successively traversed by the neutrino, each having constant electron density. The solution is then given by:
\begin{equation}\label{eq:TimeEvolution}
    \ket{\nu_\alpha(T)} = \mathcal{T}\left \{ \exp\left(-\frac{i}{\hbar} \, \int_0^T\mathcal{H}(t') \, dt'\right) \right \} \ket{\nu_\alpha(0)},
\end{equation}
where $T$ is the time the neutrino needs to cross the Earth and $\mathcal{T}$ is the time-ordering operator that ensures computations follow the correct temporal sequence of segment crossings. Note that Equation~\eqref{eq:TimeEvolution} is the matrix generalization of the solution $\exp(-i E_\nu t / \hbar) \ket{\nu(0)}$ of the Schrödinger equation with a time-independent Hamiltonian $\mathcal{H}$ in one dimension, $i \hbar \partial_t \ket{\nu(t)} = \mathcal{H} \ket{\nu(t)}$. 

After discretizing the total time $T$ into small time intervals $\Delta t_k$, the OscProb package~\citep{oscprob_v2.0.12} is used to solve Equation~\eqref{eq:TimeEvolution} numerically, by approximating $\mathcal{H}(t)$ as a sequence of piecewise-constant Hamiltonians $\mathcal{H}(t_k)$, each defined over the time interval $\Delta t_k$:
\begin{equation}\label{eq:numericalsolution}
    \ket{\nu_\alpha(T)} \approx \mathcal{T} \left \{ \prod_{k} \exp\left(-\frac{i}{\hbar} \, \mathcal{H}\left(t_k\right) \, \Delta t_k \right) \right \} \ket{\nu_\alpha(0)}.
\end{equation}
Solving Equation~\eqref{eq:numericalsolution} finally requires computing the matrix exponential of each $3 \times 3$ Hermitian matrix $\mathcal{H}(t_k)$. OscProb performs this operation efficiently by solving the eigensystem of each Hamiltonian using the hybrid algorithm described in~\citet{kopp2008efficient}, and computing the singular value decomposition of the symmetric matrix in the first term of Equation~\eqref{eq:Htotal}.

\subsection{Computation of neutrino interaction rates}
\label{subsec:nuint2}

This appendix describes the computation of neutrino interaction rates in a target volume used in this work, independent of the subsequent detection technique.

As explained in Section~\ref{subsec:nudet}, we consider eight neutrino interaction channels, which are denoted here for simplicity by $\nu_X$. For each channel $\nu_X$, the differential interaction rate with target nucleons is computed as a binned function of energy and zenith angle, per unit exposure~\citep{maderer2023unveiling}:
\begin{equation}
R^{\text{int}}_{\nu_X }(E_\nu^i, \cos\theta_z^j) \equiv \frac{\text{d}^4 N^{\text{int}}_{\nu_X }(E_\nu^i,\cos \theta_z^j)}{\text{d}E_\nu^i \, \text{d}(\cos\theta_z^j) \, \text{d}T_\mathrm{det} \, \text{d}M_\mathrm{det}},
\end{equation}
where $T_\mathrm{det}$ is the data-acquisition time, $M_\mathrm{det}$ the mass of the instrumented volume, and $E_\nu^i$ and $\cos\theta_z^j$ are the centers of the $i$-th energy bin and the $j$-th $\cos\theta_z$ bin, respectively. To compute this rate, EarthProbe weighs the atmospheric neutrino flux $\Phi_{\nu_\beta}$ by the corresponding probabilities $\text{P}_{\nu_\beta \rightarrow {\nu_X}}$ to oscillate into the flavor associated to channel ${\nu_X}$ or to stay in that flavor, and then multiplies the resulting flux arriving at the detector by the channel-dependent cross section $\sigma^{\text{int}}_{{\nu_X }}$, which is a measure of the probability that the neutrino undergoes this interaction and is given in units of area (representing the effective target size):
\begin{equation}\label{eq:interactingrate}
     R^{\text{int}}_{\nu_X} (E_\nu^i, \cos\theta_z^j) = \left( \sum_{\beta=e,\mu} \frac{\text{d}^2\Phi_{\nu_\beta}(E_\nu^i, \cos\theta_z^j)}{\text{d}E_\nu^i \, \text{d}(\cos\theta_z^j)} \, \text{P}_{\nu_\beta \rightarrow {\nu_X}} (E_\nu^i, \cos\theta_z^j) \right) \, \frac{\sigma^{\text{int}}_{\nu_X} (E_\nu^i)}{m_\text{N}}.
\end{equation}
Here, $\nu_\beta \in \{\nu_e,\nu_\mu\}$ ($\nu_\beta \in \{\bar{\nu}_e, \bar{\nu}_\mu\}$) when $\nu_X$ is a neutrino (antineutrino) channel. The contribution of atmospheric tau (anti)neutrinos is neglected, as they are very rarely produced in extensive air showers. The cross section is divided by the average mass $m_\text{N}$ of the target nucleons in the medium, so that event rates for a given detector can be obtained by multiplying $R^{\text{int}}_{\nu_X}$ by the detector mass.

Finally, for a given detector exposure $T_\mathrm{det}M_\mathrm{det}$, the expected number of neutrinos interacting within the detector in a specific energy and incidence-angle bin through channel $\nu_X$ is computed as: 
\begin{equation}\label{eq:interactingnumber}
    N^{\text{int}}_{\nu_X}(E_\nu^i,\cos\theta_z^j) = R^{\text{int}}_{\nu_X}(E_\nu^i,\cos\theta_z^j) \,\, \Delta E_\nu^i \,\, \Delta (\cos\theta_z^j) \,\, T_\mathrm{det} \, M_\mathrm{det} .
\end{equation}

\subsection{Modeling of neutrino detection}
\label{subsec:nudet2}

This appendix describes the modeling of the detector response and details the assumptions underlying the idealized detector configuration referred to in the main text as the perfect detector.

To account for detector uncertainties and inefficiencies, EarthProbe employs a set of analytical functions to model the detector response. These consist of smearing functions $\sigma(E_\nu)$ and $\sigma(\theta_z; E_\nu)$, as well as efficiency functions $\epsilon_\text{det}(E_\nu)$ and $\epsilon_\text{class}(E_\nu)$. The smearing functions encode the resolution of the reconstructed properties (note that the angular resolution is energy dependent), while the efficiency functions describe the probability of successful detection and event classification, respectively. More specifically, $\epsilon_\text{det}(E_\nu)$ represents the probability of detecting a neutrino interaction and is taken to be zero below the detector energy threshold, whereas $\epsilon_\text{class}(E_\nu)$ gives the probability of correctly assigning an interacting event to its topology (e.g., classifying a $\nu_\mu \, \mathrm{CC}$ interaction as $\nu_\mu$-like). All functions are parameterized with a small number of user-defined parameters \citep[for full details, see~][]{maderer2023unveiling}. 

The idealized next-generation detector, referred to in this work as the perfect detector, is used to assess the intrinsic sensitivity of NOTE. It thus represents the limiting case in which all detector effects are reduced to the fundamental irreducible uncertainties associated with neutrino interaction kinematics. The remaining uncertainties on the reconstructed neutrino energy and incidence angle originate solely from the stochastic nature of the interaction kinematics due to nuclear Fermi motion. In this limit, the energy resolution is negligible, while the angular resolution is parametrized as
\begin{equation}
    \sigma(\theta_z) = \frac{2.6^\circ}{\sqrt{E_\nu}}.
\end{equation}
The detector is further assumed to achieve full efficiency over the relevant energy range, and a benchmark instrumented mass of $10$~Mton is adopted.

\subsection{Linearity in the forward modeling}
\label{subsec:lin}

\begin{figure}[!htb]
    \centering
    \includegraphics[width=\linewidth]{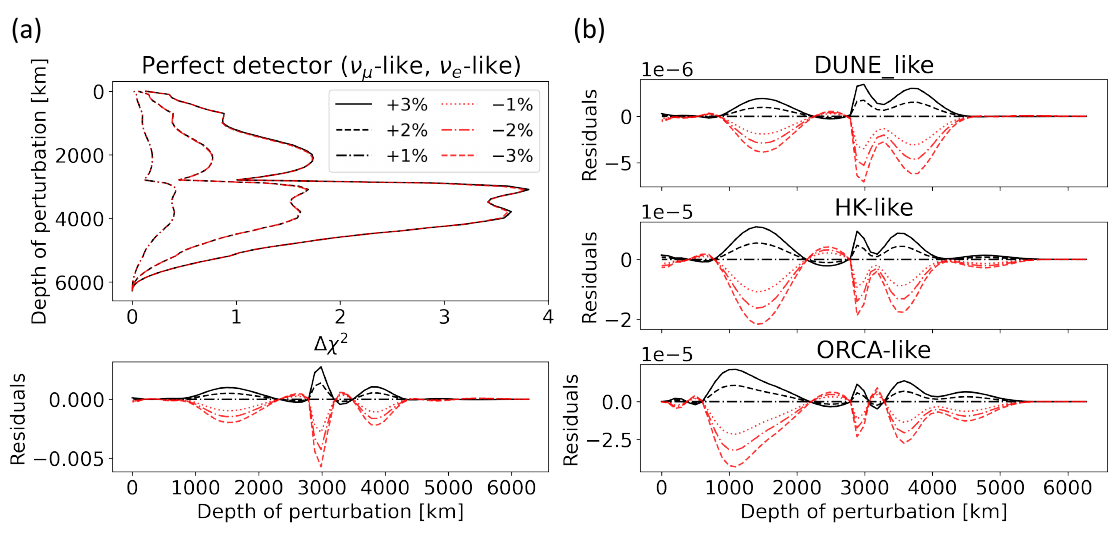}
    \caption{The upper panel in (a) shows sensitivity curves for the perfect detector (capable of distinguishing two event topologies) and for $\pm1\%$, $\pm2\%$, and $\pm3\%$ perturbations in $n_e$. The lower panel shows the corresponding residuals obtained by comparing the rescaled sensitivity curves to the $+1\%$ reference. The same residual analysis is shown in (b) for the next-generation detectors (DUNE-like, HK-like, and KM3NeT/ORCA-like, from top to bottom).}    
    \label{fig:lin}
\end{figure}

For a linearized inversion, the Fréchet derivatives must scale linearly with the amplitude of the perturbation in the model parameters. In this appendix, we demonstrate this behavior for the sensitivity curves, which differ from the Fréchet derivatives mainly through the squared difference of neutrino rates between nominal and alternative Earth models. The upper panel of Figure~\ref{fig:lin} shows sensitivity curves for the perfect detector and different perturbation amplitudes. We observe that increasing the amplitude from $1\%$ to $2\%$ ($3\%$) leads to an approximate increase by a factor of $2^2$ ($3^2$) of the sensitivity, consistent with the quadratic dependence in the $\Delta \chi^2$ definition. To quantify deviations from this approximation, the lower panel shows the residuals obtained by comparing the rescaled sensitivity curves to the $+1\%$ case. We verify that this relation also holds for realistic detectors (Figure~\ref{fig:lin}b). These results indicate that perturbations of up to $\pm3\%$ in $n_e$ remain in the linear regime for NOTE.

\bibliographystyle{plainnat} 
\bibliography{biblio}

\end{document}